\documentclass[runningheads]{llncs}
\usepackage[T1]{fontenc}
\usepackage{xcolor}
\usepackage{booktabs}
\usepackage{graphicx}
\usepackage{amsmath}
\usepackage{multirow}
\usepackage{hyperref}
\hypersetup{hidelinks}
\begin{document}
%

\title{HistDiT: A Structure-Aware Latent Conditional Diffusion Model for High-Fidelity Virtual Staining in Histopathology}
\titlerunning{HistDiT}
%
\author{Aasim Bin Saleem\inst{1} \and Amr Ahmed\inst{1} \and Ardhendu Behera\inst{1} \and Hafeezullah Amin\inst{1} \and Iman Yi Liao\inst{2} \and Mahmoud Khattab\inst{3} \and Pan Jia Wern\inst{4} \and Haslina Makmur\inst{4}}
\authorrunning{R. Aasim et al.}
%
\institute{Edge Hill University, Ormskirk, L39 4QP, United Kingdom \\
\email{\{aasimr$^*$,ahmeda,beheraa,aminh\}@edgehill.ac.uk} \and
University of Nottingham Malaysia, 43500 Semenyih, Malaysia \\ 
\email{Iman.Liao@nottingham.edu.my}, \email{Mahmoud.Khattab.Sun@gmail.com} \and
University of Southampton Malaysia, 79100 Iskandar Puteri, Malaysia \and
Cancer Research Malaysia, , 47500 Subang Jaya, Malaysia \\
\email{\{jiawern.pan,haslina.makmur\}@cancerresearch.my}}

%
\maketitle              
\begin{abstract} Immunohistochemistry (IHC) is essential for assessing specific immune biomarkers like Human Epidermal growth-factor Receptor 2 (HER2) in breast cancer. However, the traditional protocols of obtaining IHC stains are resource-intensive, time-consuming, and prone to structural damages. Virtual staining has emerged as a scalable alternative, but it faces significant challenges in preserving fine-grained cellular structures while accurately translating biochemical expressions. Current state-of-the-art methods still rely on Generative Adversarial Networks (GANs) or standard convolutional U-Net diffusion models that often struggle with ``structure and staining trade-offs''. The generated samples are either structurally relevant but blurry, or texturally realistic but have artifacts that compromise their diagnostic use. In this paper, we introduce HistDiT, a novel latent conditional Diffusion Transformer (DiT) architecture that establishes a new benchmark for visual fidelity in virtual histological staining. The novelty introduced in this work is, a) the Dual-Stream Conditioning strategy that explicitly maintains a balance between spatial constraints via VAE-encoded latents and semantic phenotype guidance via UNI embeddings; b) the multi-objective loss function that contributes to sharper images with clear morphological structure; and c) the use of the Structural Correlation Metric (SCM) to focus on the core morphological structure for precise assessment of sample quality. Consequently, our model outperforms existing baselines, as demonstrated through rigorous quantitative and qualitative evaluations. Code and trained model weights will soon be available at \href{https://github.com/AasimBinSaleem/HistDiT}{\textit{HistDiT}}.

\end{abstract}
\section{Introduction}
Breast cancer remains a significant global health concern, with over 2.3 million cases every year and nearly 0.7 million deaths, thus demanding continuous advancements in diagnostic methodologies \cite{Sedeta2023}. Precise treatments involve the evaluation of specific immune biomarkers, including estrogen/progesterone receptors (ER/PR), CD4/CD8, and Human Epidermal Growth Factor Receptor 2 (HER2) being the only FDA-approved biomarker. Accurate assessment of HER2 expression, scored at scales of $0, 1+, 2+$, or $3+$, dictates patient's eligibility for targeted therapies like Trastuzumab, significantly improving survival rates among women \cite{Golestan2024}. Over the years, Hematoxylin and Eosin (H\&E) staining serves as the cornerstone of histological analysis, but it lacks the specificity to accurately identify these biomarkers. Therefore, pathologists rely on advanced molecular techniques to obtain specialized immunohistochemical (IHC) stains. They require extensive manual processing of samples with specialized laboratory infrastructure and skilled pathologists, which is time-consuming, generates chemical waste and often prone to human errors that can alter tissue homeostasis and affect diagnostic accuracy \cite{Bai2023}. These limitations necessitate the development of digital staining alternatives for early diagnosis and treatment of breast cancer.

\textbf{Virtual staining} utilizes deep learning to generate synthetic IHC images from respective H\&E inputs, predicting specific biomarker expressions without altering the histological workflow. Several state-of-the-art virtual staining algorithms have been developed, but face significant challenges in generating accurate stains. Traditional approaches still rely on Generative Adversarial Networks (GANs), such as cCycleGAN \cite{Xu}, PyramidPix2Pix \cite{LiuBCI}, and AdaptiveSupervisedPatchNCE(ASP) \cite{Fangda}, suffer from mode collapse and often fail to capture the full diversity of pathological textures (staining). Denoising Diffusion Probabilistic Models (DDPMs) \cite{Dhariwal2021} though are promising in high-quality image generation but struggle in stain-to-stain translation due to architectural limitations. Recent works, including SynDiff \cite{Muzaffer2023}, StainDiffuser \cite{Kataria2024} and PST-Diff \cite{He2024}, primarily use Convolutional U-Nets. Although U-Net is excellent in local feature extraction but struggles to capture global context, often resulting in spatial misalignments and ``hallucinations'' in high-expression regions \cite{Kazerouni2023}. Moreover, training high-capacity diffusion models on small-scale histopathology datasets without adequate conditioning often leads to overfitting and poor generalization.

To address these limitations, we present our Histopathology Diffusion Transformer (\textbf{HistDiT}), a transformer-based architecture that integrates long-range dependencies in tissue structure and the efficient global attention mechanism for precise virtual staining. We incorporate independent conditional guidance to strictly preserve complex cellular structures and pathology-specific embeddings to provide high-level semantic context, ensuring diagnostic consistency. The core innovation of HistDiT lies in its dual-stream conditioning strategy, the multi-objective loss function and the use of Structural Correlation Metric (SCM), as briefed below:
\begin{description}
    \item[\textbf{\:a}.] We introduce a novel Diffusion Transformer (DiT) architecture that utilizes a dual-conditioning mechanism to modulate the denoising process. Unlike standard concatenation, this explicitly aligns the generated noise with both the global semantic phenotypes (extracted via UNI Model) and the spatial tissue morphology (encoded as a structural blueprint via pre-trained VAE).
    \item[\textbf{\:b}.] We propose a multi-objective robust loss function that combines an auxiliary $L_1$ term with the standard MSE to mitigate the inherent blurring effects caused by imperfect registration between serial tissue sections. This formulation effectively sharpens high-frequency details, like nucleus and membrane boundaries, that are typically smoothed out by pure Gaussian objectives.
    \item[\textbf{\:c}.] We identify that standard SSIM has a mathematical bias towards bright-field microscopy, where luminance masks structural errors. Therefore, we suggest the Structural Correlation Metric (SCM), that isolates the structure component to provide a more rigorous assessment of histological fidelity.
\end{description}

We demonstrate that HistDiT qualitatively and quantitatively outperforms state-of-the-art GAN and diffusion baselines on the BCI and MIST bench-marks. Crucially, we validate the diagnostic viability of generated IHC stains by expert pathologists via blind study, confirming superior structural fidelity compared to existing approaches.

\section{Related Work}
The task of stain-to-stain translation is defined as learning the mapping function $f:X_{H\&E} \rightarrow Y_{IHC}$. Unlike natural image translation (e.g., transforming horse-to-zebra), virtual staining requires rigorous preservation of biological structures. A generated cell nucleus must occupy the exact spatial location as in the source image; any deviation or spatial hallucination might lead to critical misdiagnosis.

\subsection{Generative Adversarial Networks (GANs) in Virtual Staining}
Early research predominantly utilized GANs \cite{Goodfellow2014} for virtually stained IHC image generation. Prominent work on H\&E to IHC translation was made by Xu et al. \cite{Xu}, where they proposed a conditional CycleGAN (cCGAN) for unpaired translation, introducing structural similarity losses patch-wise labels to maintain tissue details, but the method shows hows class-dependent inconsistencies and sensitivity to staining variations. In 2018, Isola et al. \cite{Isola} introduced Pix2Pix, a widely adopted paired translation model; however, its restrictive pixel-level losses often limit the generation of complex pathological variations. To address this, Liu et al. \cite{LiuBCI} developed PyramidPix2Pix for HER2 characterization, incorporating multi-scale processing with Gaussian filtering and introducing the BCI benchmark dataset. Overall this improved PSNR and SSIM but it struggles with high-expression regions (level 3+) being washed out. Recently, Duan et al. \cite{Duan} separated structural content and staining style using attention mechanisms, but their reliance on PatchGAN that assumes pixel-level independence among patches, lead to global staining inconsistencies. Ultimately, GANs remain difficult to train and prone to mode collapse, thus limiting the diversity in image generation and their clinical usage \cite{Ainur2023}.

\subsection{Diffusion Models in Stain-to-Stain Translation}
Denoising Diffusion Probabilistic Models (DDPMs) \cite{Ho2020} have recently surpassed GANs in high-resolution image synthesis and stain translation by iteratively refining noisy images. Latent Diffusion Models (LDMs) further improved efficiency by operating in a compressed latent space while maintaining high visual quality \cite{Rombach2022}. In histopathology for cancer diagnosis, DDPMs are emerging as a valuable tool for generating diverse synthetic data. Moghadam et al. \cite{Moghadam2023} first explored that diffusion models could generate diverse tissue textures, avoiding GAN-based instability. Later, Kataria et al. \cite{Kataria2024} proposed \textit{StainDiffuser}, a dual-diffusion architecture that simultaneously performs cell-specific segmentation and IHC staining on H\&E images. While effective for specific markers (CK8/18, CD3) but study highlights a critical gap between quantitative metrics and diagnostic use. Xuanhe et al. \cite{Xuanhe2024} proposed conditional DDPM for HER2 expression levels assessment, utilizing an attention U-Net conditioned on H\&E with classifier-free guidance (CFG). Reported better image quality metrics while operating at $64\times64$. He et al. \cite{He2024} also introduced \textit{PST-Diff}, a latent diffusion method for HER2 assessment using asymmetric attention, dynamic variance scheduling and latent-transfer modules. Improving metrics over GANs but failing to match PyramidPix2Pix in structural similarity (SSIM) on BCI benchmark.

To address spatial misalignment, Liu et al. \cite{StarDiff2025} proposed \textit{Star-Diff}, combining deterministic restoration path to preserve structure and stochastic path for molecular variability. Similarly, Großkopf et al. introduced \textit{HistDiST} \cite{HistDiST}, which utilizes a DDIM sampler with a cosine schedule to balance structure and diversity. However, \textit{HistDiST} reported lower SSIM scores for HER2, suggesting that the accelerated, deterministic sampling of DDIM may compromise the reconstruction of high-frequency membrane textures (essential for grading) as the model has fewer opportunities to correct alignment errors. These findings indicate that while diffusion models offer superior diversity, but existing CNN-based architectures struggle with the precision required for clinical diagnostics.

\subsection{Diffusion Transformers (DiTs)}
While GANs and DDPMs traditionally rely on CNN-based backbones like U-Nets, recent advancements suggest that Transformer backbones can outperform CNNs in capturing high-resolution representations and long-range dependencies. State-of-the-art models like GPT-4 \cite{OpenAI2023}, DALL-E 3 \cite{Lin2023}, and Stable Diffusion 3.5 \cite{Stability2024} have successfully integrated Transformers. Peebles et al. \cite{Peebles2023} formally introduced Diffusion Transformers (DiTs), treating diffusion as a sequence-to-sequence problem. Unlike U-Nets, which are biased towards local processing, DiTs use global attention and Adaptive Layer Normalization (adaLN) to integrate conditioning information effectively \cite{Wu2024}. Our proposed HistDiT adapts this scalable architecture for virtual staining by extending the standard DiT conditioning mechanism to support both spatially dense inputs (the H\&E structural map) and semantically rich embeddings (the UNI phenotype vector) simultaneously, a novel configuration in the context of medical image synthesis.

\subsection{Foundation Models as Clinical Priors}
Foundation models like UNI \cite{Chen2024} and CTransPath \cite{Xiyue2024}, trained on massive whole-slide image (WSI) datasets using self-supervised learning, have revolutionized feature extraction. UNI \cite{Chen2024}, a ViT-H model, captures high-level semantic concepts—such as tumor grade and lymphocytic infiltration—that are invariant to local deformations. In HistDiT, we utilize UNI as a ``semantic prior provider'', injecting diagnostic context into the generative process to help model easily distinguishing cellular structures (e.g. lymphocytes from tumor nuclei) that low-level pixel data alone cannot resolve.

\section{Proposed Methodology}
\begin{figure}
\includegraphics[width=\textwidth]{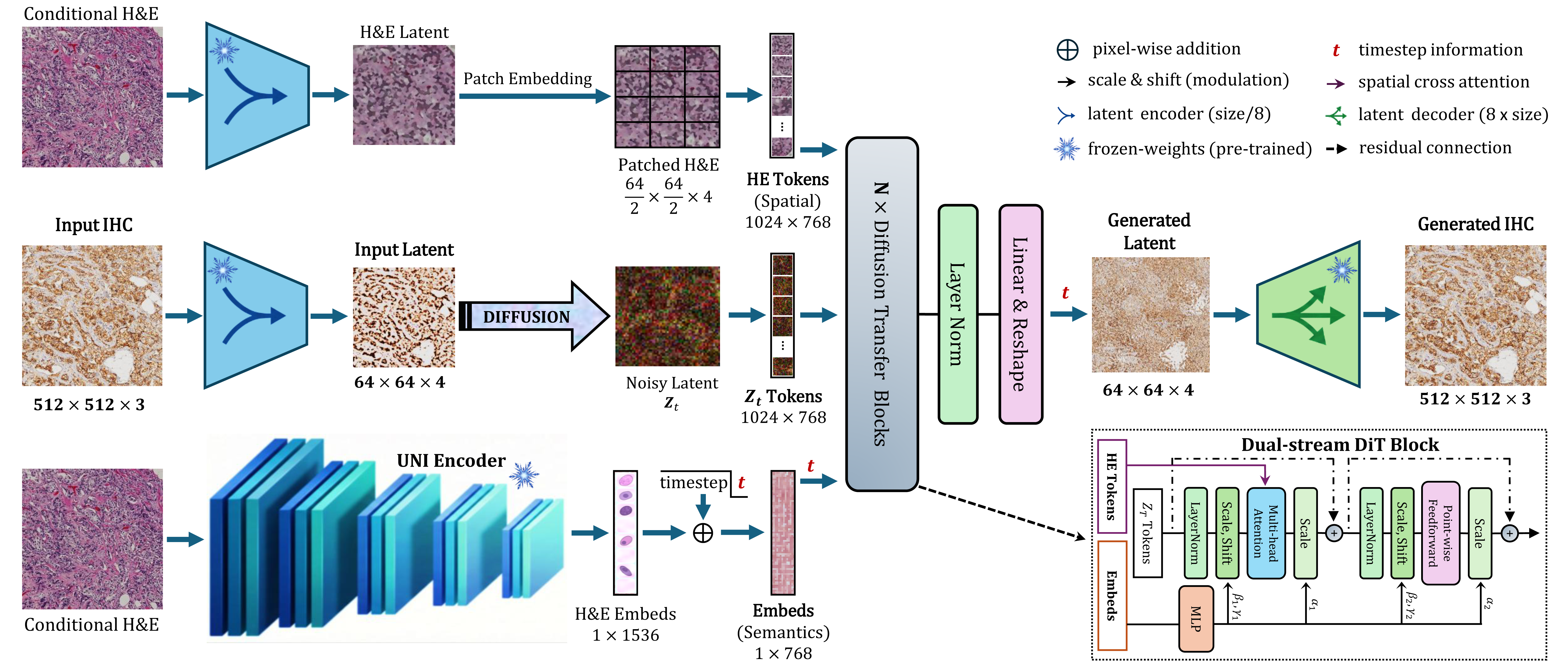}
\caption{\textbf{The HistDiT Architecture} replaces the standard U-Net with a Transformer backbone that integrates two distinct conditioning streams: (i) \textbf{Global Semantic Guidance} uses frozen UNI embeddings injected via Adaptive Layer Norm (adaLN) to enforce diagnostic consistency; (ii) \textbf{Spatial Structural Guidance} uses VAE-encoded H\&E latents injected via Cross-Attention to strictly preserve tissue morphology.} \label{fig1}
\end{figure}

This paper introduces our \textbf{HistDiT (Histopathology Diffsuion Transformer)}, a novel virtual staining method for generating realistic and accurate IHC stained images used for HER2 assessment in breast cancer diagnostics. HistDiT utilises a dual-conditioned Diffusion Transformer to translate H\&E images into specialized IHC stains with high structural and semantic fidelity. The overall framework is illustrated in Fig.~\ref{fig1}.
\noindent We construct our approach upon conditional DDPMs \cite{Nichol2021}, which learn to approximate a data distribution $q(x_0)$ by reversing a gradual noising process. The \textbf{forward process} $q(x_t|x_0)$ iteratively adds Gaussian noise $\epsilon \sim \mathcal{N}(0, \mathbf{I})$ according to a variance schedule $\beta_t$, such that a sample at timestep $t$ can be expressed in closed form as $x_t = \sqrt{\bar{\alpha}_t} x_0 + \sqrt{1 - \bar{\alpha}_t} \epsilon$, where $\bar{\alpha}_t$ is the cumulative signal after noise addition. The \textbf{reverse process} $p_\theta(x_{t-1}|x_t)$ is then parametrized as learned Gaussian shifts where a neural network $\epsilon_\theta(x_t, t, c)$ predicts the noise component to denoise $x_t$ conditioned on context $\textbf{c}$. The model is trained to minimize the variational lower bound, simplified to the mean squared error (MSE) between the true noise $\epsilon$ and the predicted noise $\epsilon_\theta$: $\mathcal{L}_{MSE} =  \textbf{E}_{x_0, t, \epsilon} \left( \| \epsilon - \epsilon_\theta(x_t, t, \textbf{c}) \|_2^2 \right)$.

\subsection{Latent Diffusion with HistDiT}
To reduce computational complexity while maintaining high-resolution fidelity, we operate in the compressed latent space of a pre-trained Variational Auto-Encoder (VAE). We utilize the Stable Diffusion VAE ($\mathcal{E}, \mathcal{D}$), which compresses the input $I \in \mathbf{R}^{H \times W \times 3}$ into a latent representation $z = \mathcal{E}(I) \in \mathbf{R}^{\frac{H}{8} \times \frac{W}{8} \times 4}$ \cite{StabilityVAE}. Diffusion is performed on the latent vector $z$, and the final image is reconstructed via $\tilde{I} = \mathcal{D}(z)$. To align the latent distribution with the standard normal prior expected by the diffusion model, we normalize the latent space to unit variance following standard latent diffusion protocols \cite{Rombach2022}.
\subsubsection{Dual-Stream Transformer Architecture.}
Unlike traditional U-Net backbones that rely on simple channel concatenation (which often dilutes structural guidance), HistDiT strategically handles two distinct streams of conditioning: Semantic and Spatial. This separation is critical for virtual staining, where the model must alter the biochemical appearance (stain intensity) without hallucinating or distorting the physical tissue structure. The noisy latent $z_t$ is first ``patchified'' into a sequence of $N$ tokens, where $N = (h/p) \times (w/p)$ and $p$ is the patch size ($p=2$). The DiT block consists of standard multi-head self-attention (MHSA), layer normalization (LN), and pointwise feedforward (FF) layers, but innovates in how conditions are injected (Fig.~\ref{fig1}: flagged branches).

\noindent \textbf{i. Global Semantic Guidance via adaLN:}
To resolve diagnostic ambiguity (e.g., differentiating HER2 level 1+ from 2+ based on subtle stain intensity), we utilize the ``UNI Foundation Model'' as a semantic prior \cite{Chen2024}. The input H\&E image $x_H$ is processed by the frozen UNI encoder to extract a global phenotype embedding $c_{sem} \in \textbf{R}^{1536}$. This vector added with the timestep embedding $t_{emb}$ is injected into transformer block using Adaptive Layer Normalization (adaLN)\cite{Peebles2023}. Specifically, a simple MLP regresses $\textit{MLP}(c_{combined})$ into dimension-wise scale ($\gamma$) and shift ($\beta$) parameters, modulating the normalized latent features $Z_t$ as:
\begin{equation}
    \text{adaLN}(Z_t, c_{combined}) = \gamma(c_{combined}) \odot \textit{LN}(Z_t) + \beta(c_{combined})
\end{equation}
Here, $\odot$ denotes element-wise multiplication. This allows high-level pathological concepts (like tumor grade, tissue subtype, biomarker information) to globally influence the generative statistics (the style and intensity) without disrupting local structure \cite{Peebles2023}. This ensures that the biochemical expression levels are consistent with the tissue phenotype.

\noindent \textbf{ii. Spatial Structural Guidance via Cross-Attention:}
To ensure the generated IHC image perfectly respects the cellular morphology of the source H\&E, we encode the H\&E image $x_H$ into a spatial latent map $c_{spatial} = \mathcal{E}(x_H) \in \textbf{R}^{h \times w \times 4}$ which is then flattened into a sequence of spatial tokens. This structural blueprint is injected using Multi-Head Cross-Attention layers within the DiT blocks \cite{Alexey2020}. The intermediate representation of noisy latent $Z_t$ serves as the Query ($Q$), while $c_{spatial}$ acts as the Key ($K$) and Value ($V$). This mechanism forces the model to attend to specific spatial locations in the H\&E blueprint when synthesizing high-frequency details (e.g., nuclei). Unlike simple concatenation, cross-attention allows the model to dynamically attend to relevant structural features at different stages of the denoising process, thereby preventing the hallucinations.

\noindent \textbf{iii. Classifier-Free Guidance:}
We utilize Classifier-Free Guidance (CFG)  to include adherence to this dual-conditioning constraints. During training, both semantic ($c_{sem}$) and spatial ($c_{spatial}$) conditions are randomly dropped with probability $p_{drop} = 0.11$ to learn an unconditional prior. During inference, the noise prediction is extrapolated to amplify the conditional signal. Defined as,
\begin{equation}
    \tilde{\epsilon}_\theta = \epsilon_\theta(z_t, \emptyset) + scale \cdot \left(\epsilon_\theta(z_t, c_{sem}, c_{spatial}) - \epsilon_\theta(z_t, \emptyset)\right)
\end{equation}
We empirically set the guidance scale to $3.0$. This pushes the generation towards high diagnostic fidelity without sacrificing the staining diversity \cite{Nichol2021}.

\subsection{Multi-Objective Loss Function}
Standard diffusion models minimize the MSE between the predicted and actual noise. However, histopathology datasets use serial sections that introduce inevitable spatial misalignment. Pure MSE heavily penalizes slight pixel shifts, causing the model to produce blurry ``average'' images to minimize variance. To address this, we introduce an auxiliary $L_1$ objective on the noise prediction. The $L_1$ norm is less sensitive to outliers (misaligned edges) and encourages sparsity in the error, resulting in sharper boundaries. The total loss $\mathcal{L}_{total}$ is:
\begin{equation}
    \mathcal{L}_{total} = \lambda_{MSE} . \textbf{E}_{z_0, \epsilon, t} \left[ \| \epsilon - \epsilon_\theta(z_t, t, c) \|_2^2 \right] + \lambda_{L1} \textbf{E}_{z_0, \epsilon, t} \left[ \| \epsilon - \epsilon_\theta(z_t, t, c) \|_1 \right]
\end{equation}
Empirically, we set $\lambda_{MSE} = 0.7$ and $\lambda_{L1} = 0.3$. This hybrid loss is a critical contributor to the model's ability to generate sharp cellular membranes despite imperfect ground truth alignments, as shown in Fig.~\ref{fig:objective_ablation}.

\section{Experiments and Evaluation}
We evaluate HistDiT on the benchmark BCI dataset \cite{LiuBCI} and the Multi-IHC Stain Translation (MIST) dataset \cite{Fangda}, publicly available collection of paired H\&E and IHC images specifically designed for histopathological image translation on HER2 assessment. Each dataset comprises approximately 5,000 paired patches ($1024 \times 1024$) sourced from several WSIs. They cover all clinically relevant HER2 expression levels (0, 1+, 2+, 3+). The datasets are split between 4k pairs for training and 1k for testing, consistent with standard split used in prior works.

\subsection{Implementation Details}
Due to computational constraints and the patch size of DiT, images were resized to $512 \times 512$ and normalized to $[-1, 1]$. We used the DiT-B/2 backbone configuration \cite{Peebles2023}. This choice was empirically determined to balance computational efficiency (Base model) with the high spatial resolution (patch size of 2) required to resolve fine-grained histological structures like nuclei boundaries. We utilized the pre-trained VAE ($sd\text{-}vae\text{-}ft\text{-}mse$) to compress inputs into the latent space, applying the standard scaling factor to ensure unit variance \cite{Rombach2022}. The UNI model (UNI2-h) is used in a frozen state to extract 1536-dimensional embeddings. The model was trained on an NVIDIA RTX 6000 (48GB) GPU using the AdamW optimizer with a learning rate of $3 \times 10^{-5}$ and a batch size of 8 for 1000 epochs. Notably, we use the scaled-Linear Noise Schedule ($\beta_{start}=0.0001, \beta_{end}=0.02$); that prevents rapid drop in signal-to-noise ratio (SNR) during early timesteps, critical for preserving key morphological features in histopathology.

\subsection{Quantitative Metrics}
We use standard metrics including Mean Square Error (MSE), Peak Signal-to-Noise Ratio (PSNR) \cite{PSNR2010}, Structural Similarity Index (SSIM) \cite{SSIM2004}, Learned Perceptual Image Patch Similarity (LPIPS) \cite{Zhang2018}, and vanilla Fréchet Inception Distance (FID) \cite{Parmar2022}. Additionally, to address the limitations of standard SSIM in histopathology, we report the Structural Correlation Metric (SCM). The standard SSIM is heavily biased by the white background (high intensity values), mathematically inconsistent for Virtual Staining because the high luminance of the white background dominates the score, masking structural errors. SSIM is a product of luminance, contrast and structure. For a generated IHC $y'$ to be compared with real IHC $y$, it is,
$$ SSIM(y,y') = f(l(y,y'), c(y,y'), s(y,y'))$$
\begin{equation}
    SSIM(y,y') = \frac{2\mu_y\mu_y' + C_1}{\mu_y^2 + \mu_{y'}^2 + C_1} . \frac{2\sigma_{yy'} + C_2}{\sigma_y^2 + \sigma_{y'}^2 + C_2} . \frac{\sigma_{yy'} + C_3}{\sigma_y \sigma_y' + C_3}
\end{equation}
The luminance term compares mean brightness; however, a generated image dominated by bright pixels (even with poor cellular structures) will often make a mean brightness ($\mu$) close to the ground truth. This results in a luminance score near 1.0, which acts as a multiplier that bloats the final SSIM value while masking a poor structural score. In histopathology, where preserving tiny, complex structures is critical, evaluation metrics must prioritize structural correlation. Therefore, we isolate structure ($s$) component to provide a honest assessment.

\subsubsection{The Structural Correlation Metric (SCM)} is a split of the structure component of multi-scale SSIM ~\cite{MS-SSIM} with a window-size of $11\times11$, given by,

\begin{equation}
SCM(y, y') = \frac{1}{MN} \sum_{i=1}^{M} \sum_{j=1}^{N} \left(\frac{\sigma_{yy'}(i,j) + C}{\sigma_{y}(i,j) \; \sigma_{y'}(i,j) + C}\right)
\end{equation}
This metric focuses purely on the correlation of variance (texture and edges), ignoring mean luminance shifts inherent in staining differences.

\section{Results and Discussion}
We compared HistDiT against established GAN and diffusion baselines on the MIST and BCI test datasets. Our method presents quality staining and successfully achieves superior performance across perceptual and structural metrics. 

\subsection{Experimental Results on BCI Benchmark}
\begin{figure*}[htbp]
\centering
\setlength{\tabcolsep}{0.5pt} 
\renewcommand{\arraystretch}{0.5} 
\definecolor{levelorange}{RGB}{210,105,30}
\definecolor{brown(web)}{rgb}{0.65, 0.16, 0.16}
\resizebox{\textwidth}{!}{%
\begin{tabular}{c c c c c c c c c}
\rotatebox{90}{\parbox{1.3cm}{\centering \textbf{H\&E}}} & 
\includegraphics[width=0.115\linewidth]{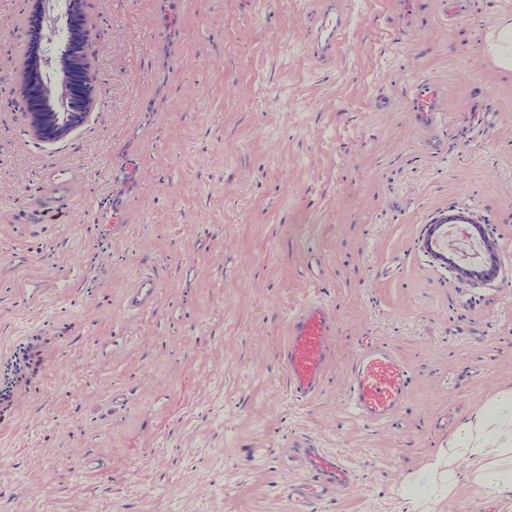} & 
\includegraphics[width=0.115\linewidth]{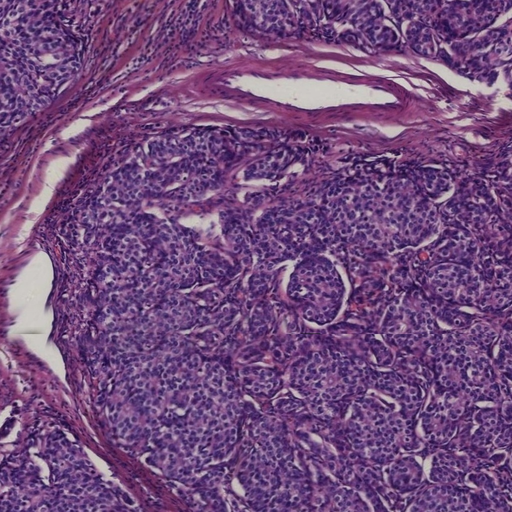} & 
\includegraphics[width=0.115\linewidth]{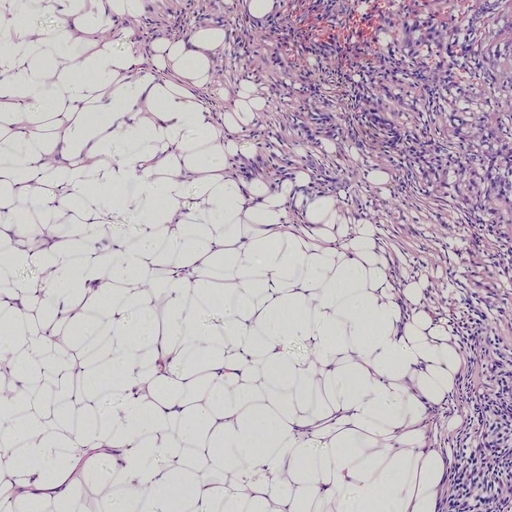} & 
\includegraphics[width=0.115\linewidth]{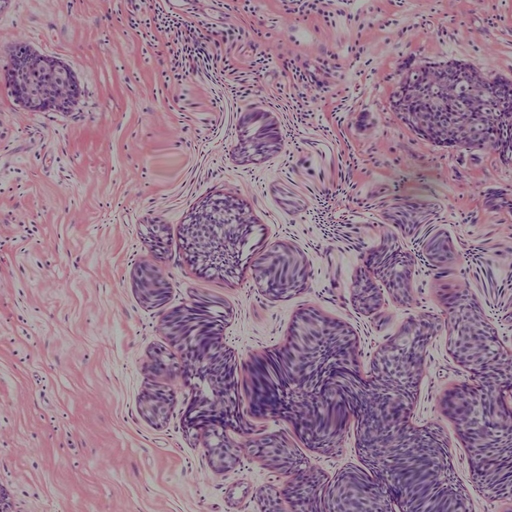} & 
\includegraphics[width=0.115\linewidth]{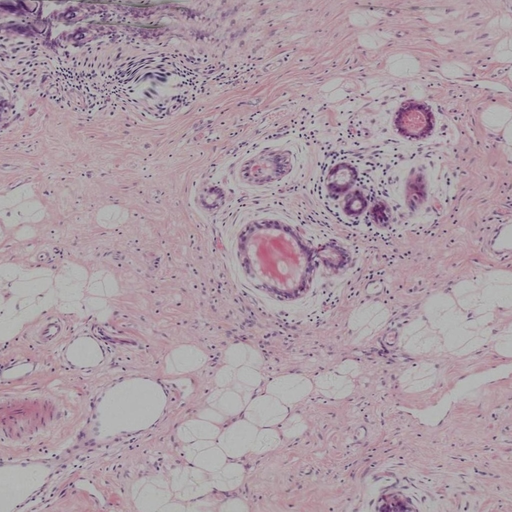} & 
\includegraphics[width=0.115\linewidth]{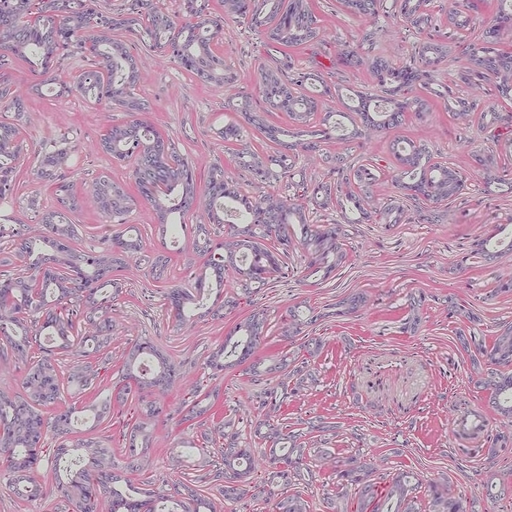} & 
\includegraphics[width=0.115\linewidth]{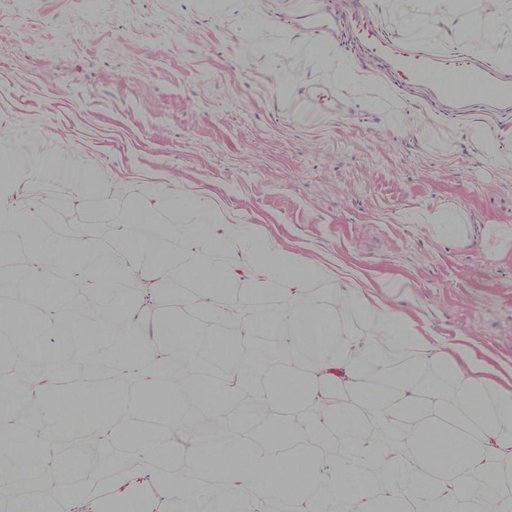} & 
\includegraphics[width=0.115\linewidth]{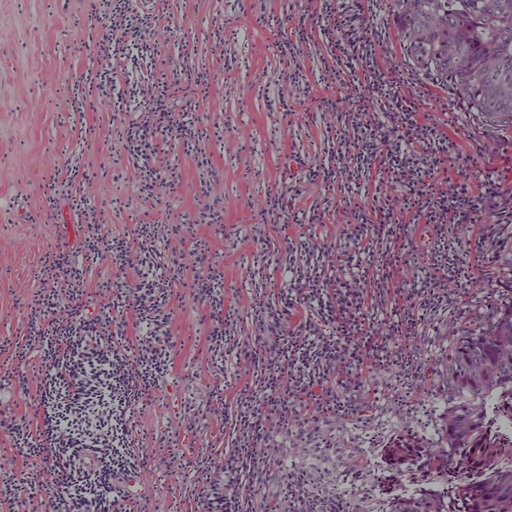} \\

\rotatebox{90}{\parbox{1.3cm}{\centering \textbf{cDiff}\cite{Xuanhe2024}}} & 
\includegraphics[width=0.115\linewidth]{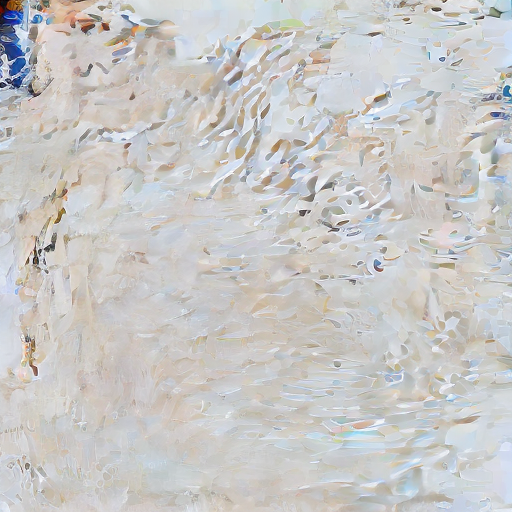} & 
\includegraphics[width=0.115\linewidth]{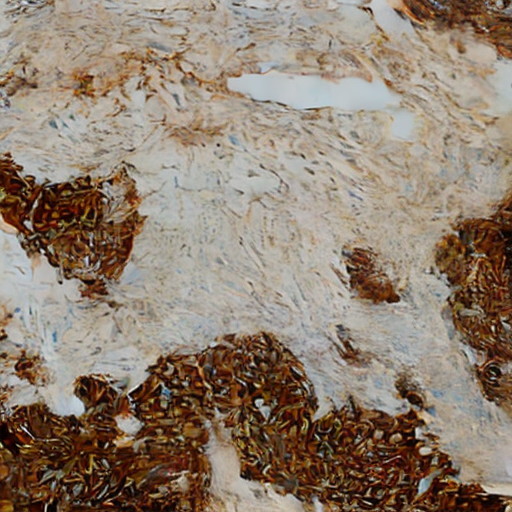} & 
\includegraphics[width=0.115\linewidth]{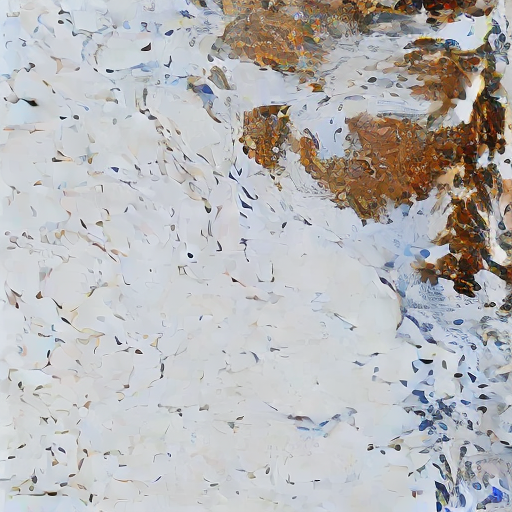} & 
\includegraphics[width=0.115\linewidth]{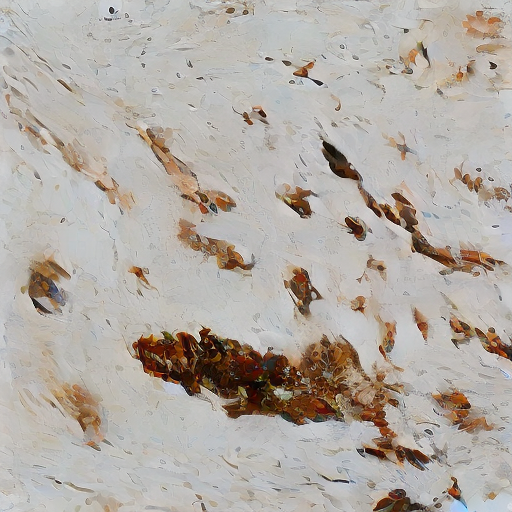} & 
\includegraphics[width=0.115\linewidth]{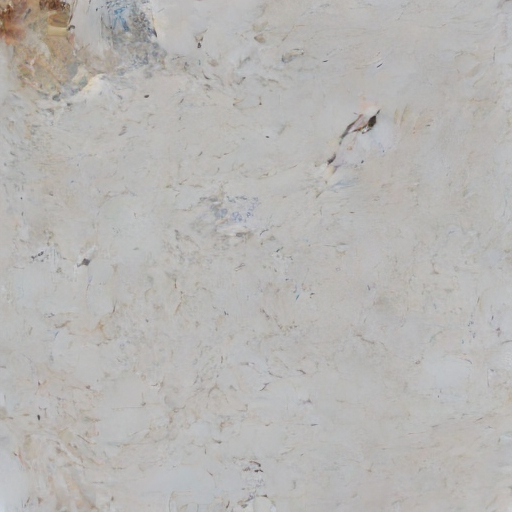} & 
\includegraphics[width=0.115\linewidth]{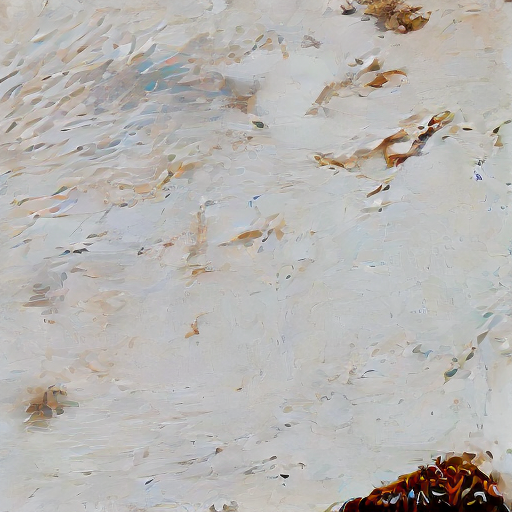} & 
\includegraphics[width=0.115\linewidth]{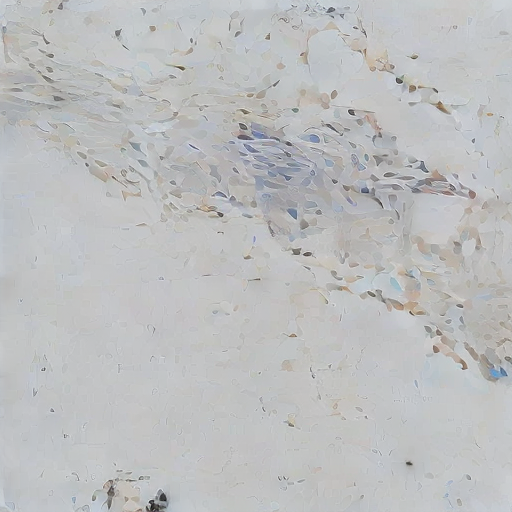} & 
\includegraphics[width=0.115\linewidth]{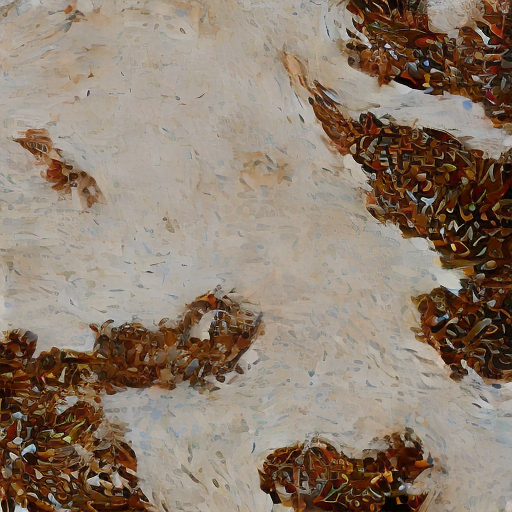} \\

\rotatebox{90}{\parbox{1.3cm}{\centering \textbf{pP2P}\cite{LiuBCI}}} & 
\includegraphics[width=0.115\linewidth]{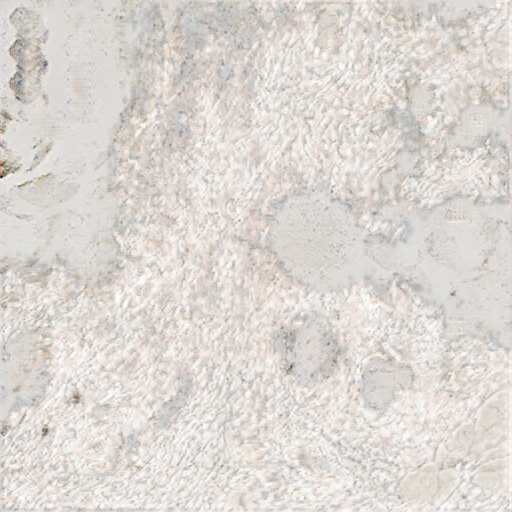} & 
\includegraphics[width=0.115\linewidth]{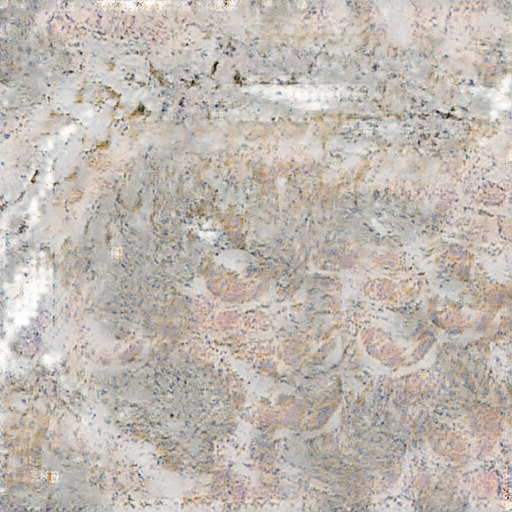} & 
\includegraphics[width=0.115\linewidth]{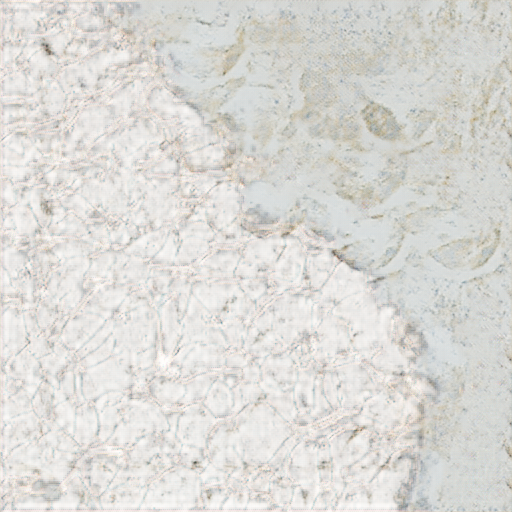} & 
\includegraphics[width=0.115\linewidth]{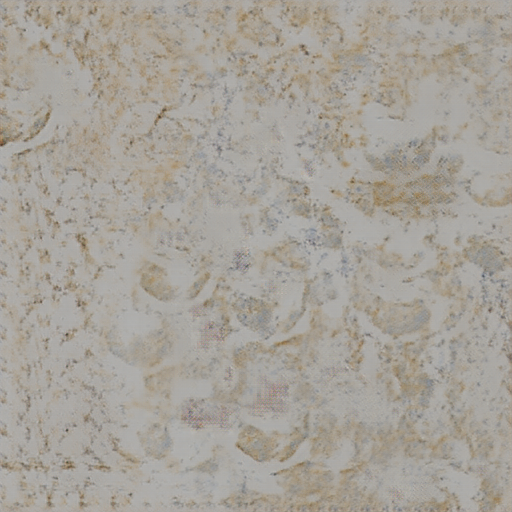} & 
\includegraphics[width=0.115\linewidth]{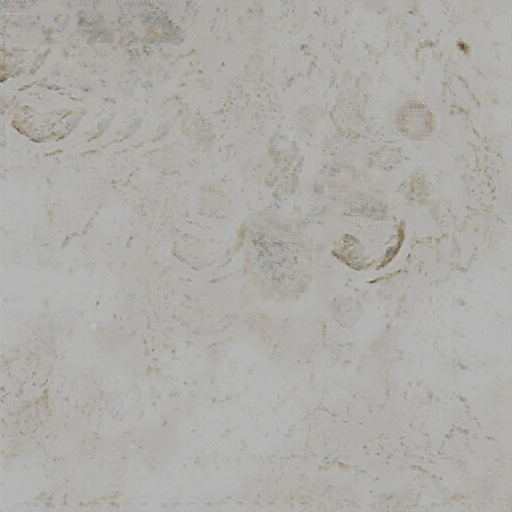} & 
\includegraphics[width=0.115\linewidth]{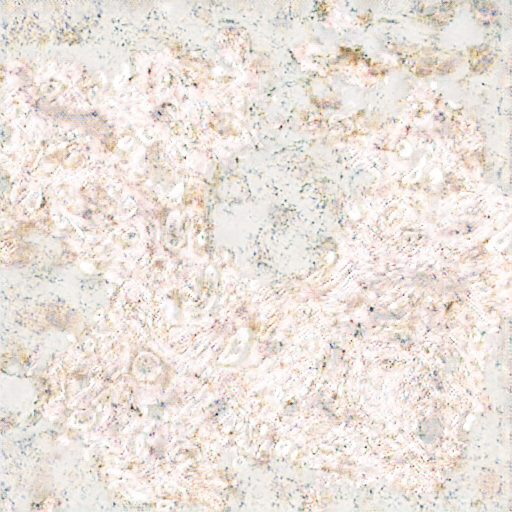} & 
\includegraphics[width=0.115\linewidth]{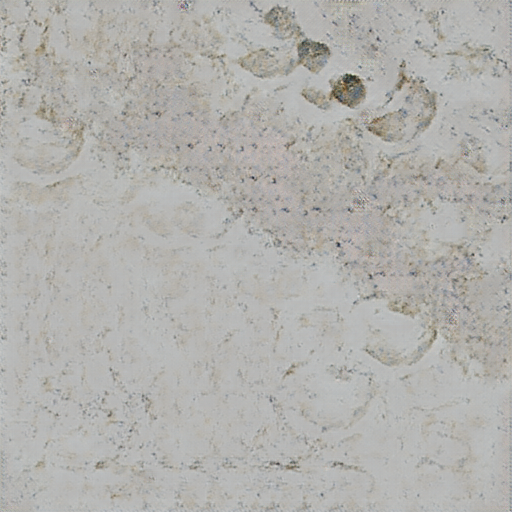} & 
\includegraphics[width=0.115\linewidth]{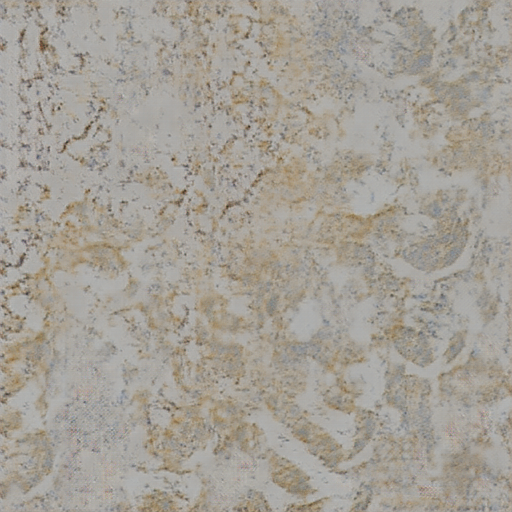} \\

\rotatebox{90}{\parbox{1.3cm}{\centering \textbf{Ours}}} & 
\includegraphics[width=0.115\linewidth]{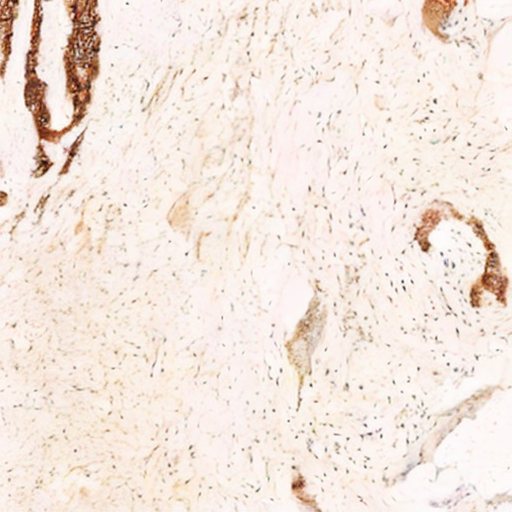} & 
\includegraphics[width=0.115\linewidth]{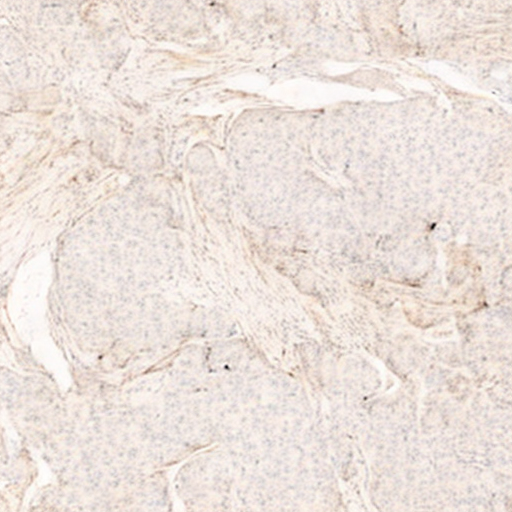} & 
\includegraphics[width=0.115\linewidth]{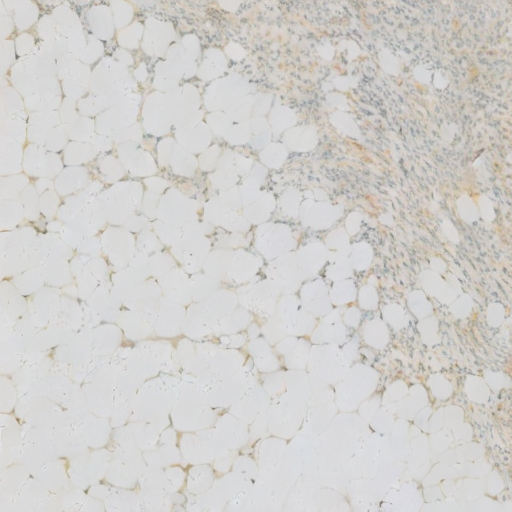} & 
\includegraphics[width=0.115\linewidth]{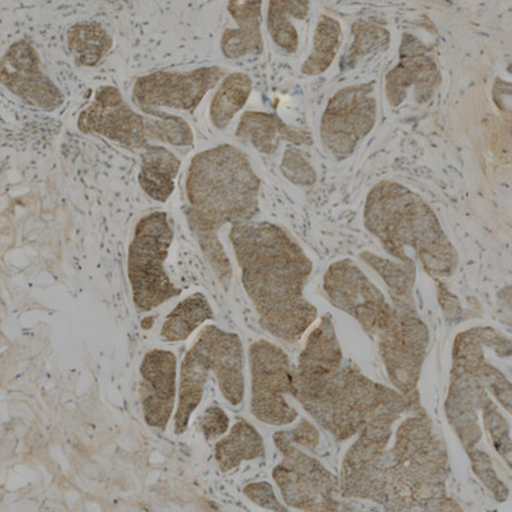} & 
\includegraphics[width=0.115\linewidth]{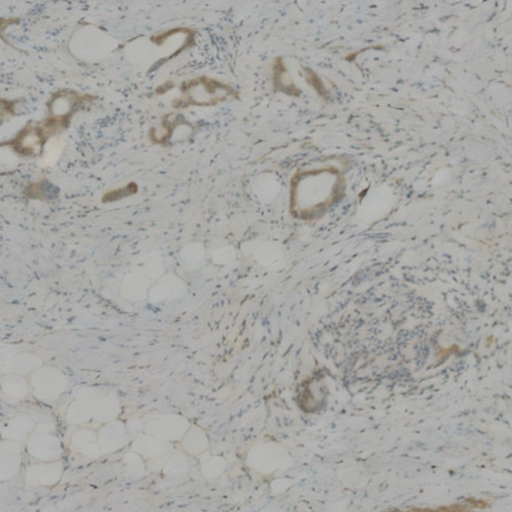} & 
\includegraphics[width=0.115\linewidth]{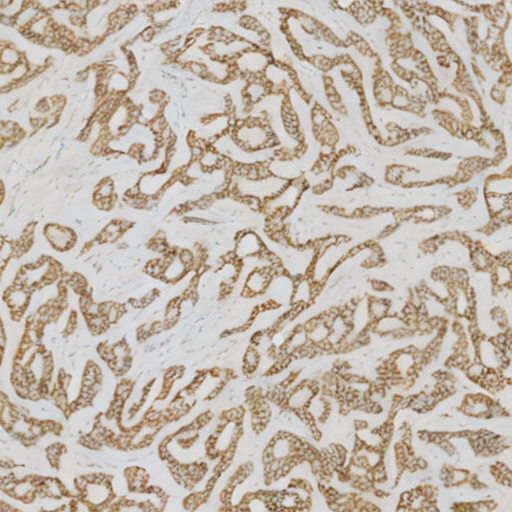} & 
\includegraphics[width=0.115\linewidth]{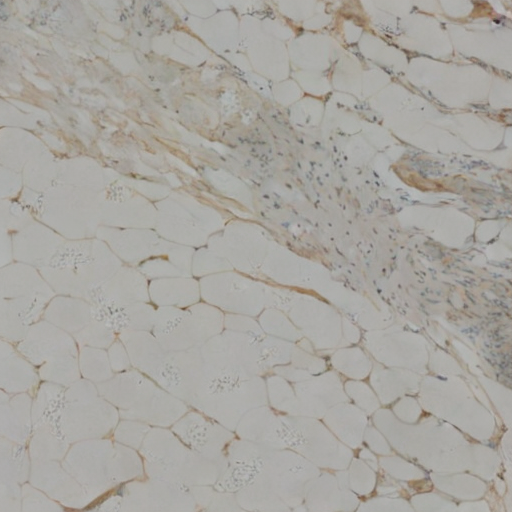} & 
\includegraphics[width=0.115\linewidth]{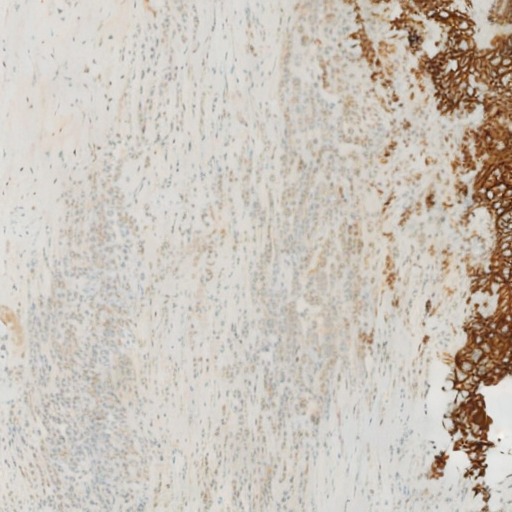} \\
\rotatebox{90}{\parbox{1.3cm}{\centering GT \textbf{IHC}}} & 
\includegraphics[width=0.115\linewidth]{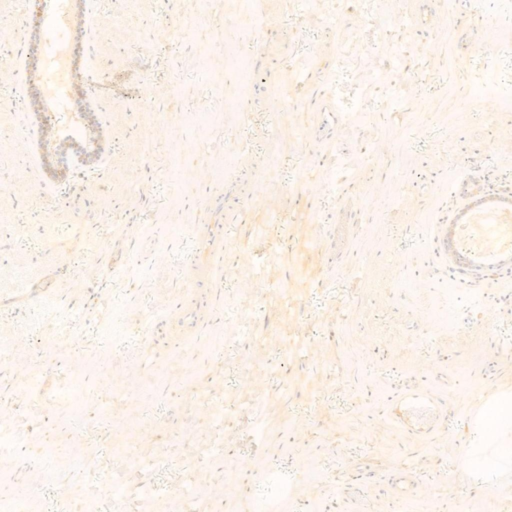} & 
\includegraphics[width=0.115\linewidth]{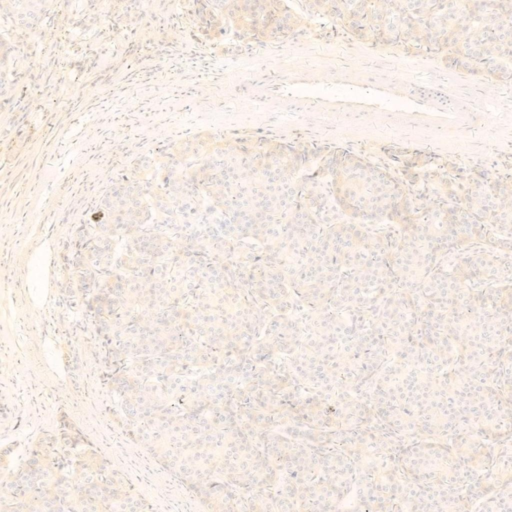} & 
\includegraphics[width=0.115\linewidth]{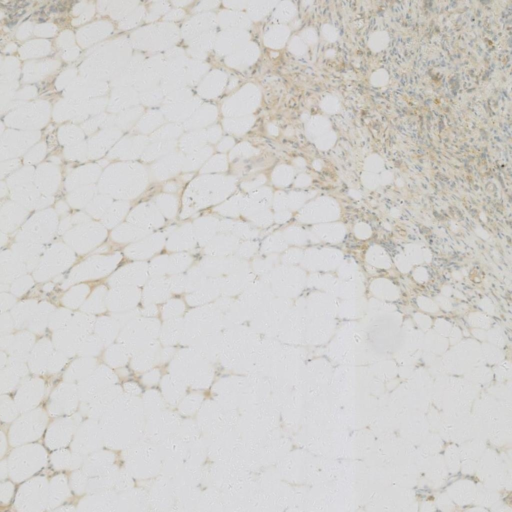} & 
\includegraphics[width=0.115\linewidth]{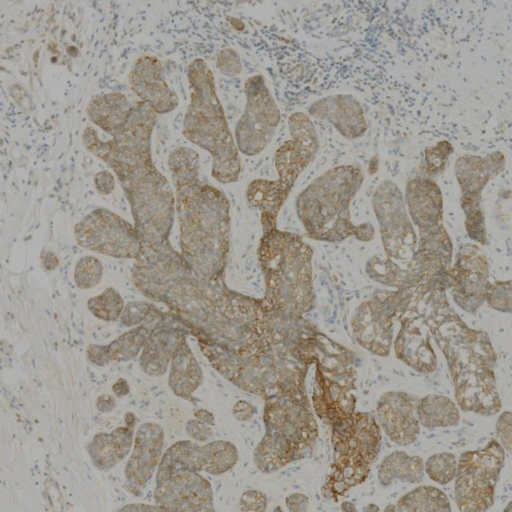} & 
\includegraphics[width=0.115\linewidth]{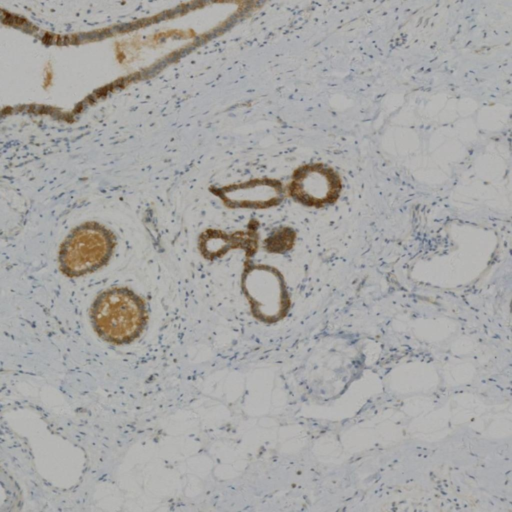} & 
\includegraphics[width=0.115\linewidth]{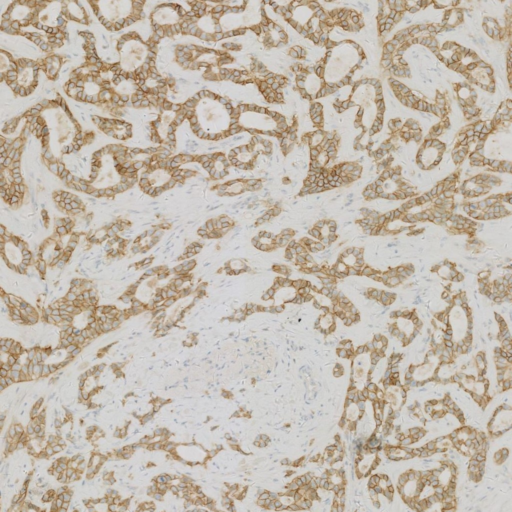} & 
\includegraphics[width=0.115\linewidth]{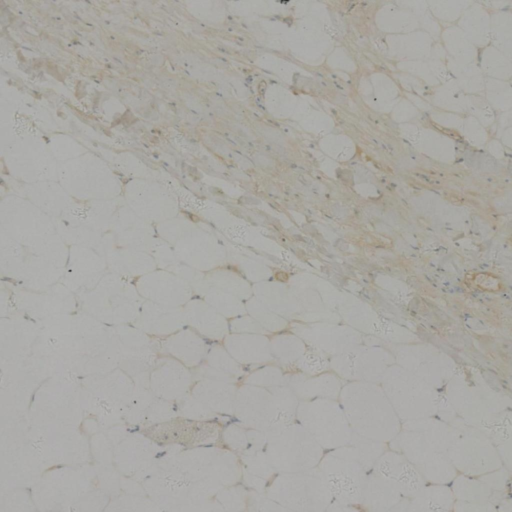} & 
\includegraphics[width=0.115\linewidth]{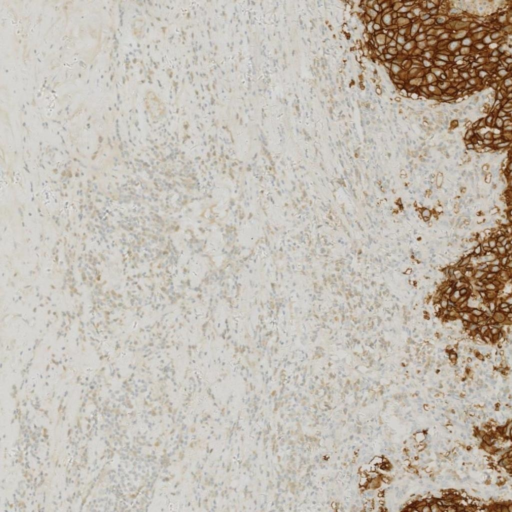} \\

 & \centering
\textcolor{orange}{Level 0} & 
\textcolor{levelorange}{Level 1+} & 
\textcolor{levelorange}{Level 1+} & 
\textcolor{brown(web)}{Level 2+} & 
\textcolor{brown(web)}{Level 2+} & 
\textcolor{red}{Level 3+} & 
\textcolor{red}{Level 3+} & 
\textcolor{red}{Level 3+} \\

\end{tabular} 
}
\caption{Qualitative comparison on BCI Dataset across HER2 expression levels, ranging from negative (Level 0) to strongly positive (Level 3+). Rows compare the input H\&E and baseline methods against our HistDiT and Ground Truths. HistDiT demonstrates higher fidelity and accurate stain intensity, particularly in high-grade regions (2+, 3+).}
\label{fig:visual_table}
\end{figure*}

The BCI dataset serves as a rigorous benchmark for evaluating breast cancer due to its complex tissue morphology and subtle HER2 staining variations. It demands high fidelity in preserving pathological details, critical for accurate diagnosis. Test data contains 977 paired H\&E, IHC samples across varying HER-2 expression levels. A qualitative comparison of the generated samples, with state-of-the-art algorithms, is shown in Fig. ~\ref{fig:visual_table}, highlighting the model's superior performance in handling complex HER-2 expression patterns.
Visual results demonstrate that HistDiT effectively distinguishes between subtle staining variations, particularly in high-expression (Level 3+) samples where baselines often produce ``washed out'' artifacts. We observed few cases where generated samples diverge from Ground Truth. This is attributed to serial sectioning artifacts inherent to the dataset, where the Input H\&E and Ground Truth IHC are physically different tissue slices cut $4\text{--}5 \mu m$ apart. Due to this depth disparity, cellular structures visible in the H\&E may terminate or shift position in the subsequent IHC slice. In these instances, HistDiT strictly adheres to the morphological structures present in the source H\&E for cellular staining, prioritizing diagnostic safety over hallucination.
\begin{table}[htbp]
\caption{Quantitative comparison with State-of-the-Art methods on BCI dataset. The best are highlighted in \textbf{bold}, and the second-best are \underline{underlined}. Symbol ${\rho}$ indicates pixel space operation and $z$ means latent space. ($\downarrow$ lower is better, $\uparrow$ higher is better)}
\label{tab:BCIcomparison}
\centering
\resizebox{\textwidth}{!}{%
\begin{tabular}{l|c|c|c|c|c|c}
\hline
\textbf{Method/Model} & \textbf{MSE}$\downarrow$ & \textbf{PSNR(dB)}$\uparrow$ & \textbf{SSIM}$\uparrow$ & \textbf{SCM}$\uparrow$ & \textbf{LPIPS}$\downarrow$ & \textbf{FID}$\downarrow$ \\ \hline
Cycle GAN \cite{Xu}$^{\rho}$ & 1892.00 & 15.98 & 0.372 & 0.447 & 0.624  & 188.33 \\
Pix2Pix \cite{Isola}$^{\rho}$ & 1560.26 & 17.08 & 0.303 & 0.391 & 0.769  & \underline{160.39} \\
Pyramid Pix2Pix \cite{LiuBCI}$^{\rho}$ & \underline{1348.54} & 19.61 & 0.397 & 0.473 & \underline{0.466} & 167.40 \\
ASP \cite{Fangda} $^{\rho}$ & --- & --- & \underline{0.5032} & --- & --- & --- \\
Syn Diff \cite{Dhariwal2021}$^{z}$ & --- & 14.28$\pm$2.52 & 0.32$\pm$0.1 & --- & --- & --- \\
PST-Diff \cite{Muzaffer2023}$^{z}$ & --- & 16.75$\pm$4.20 & 0.38$\pm$0.1 & --- & --- & --- \\
Conditional Diff. \cite{Xuanhe2024}$^{z}$ & 3011.36 & 15.76 & 0.4050 & \underline{0.494} & 0.591 & 194.82 \\ 
Star-Diff (No Visual)\cite{StarDiff2025}$^{z}$ & --- & \underline{21.30$\pm$0.01} & 0.5301 & --- & --- &--- \\
HistDiST \cite{HistDiST}$^{z}$ & --- & --- & 0.4693 & --- & ---& --- \\ \hline
Proposed [HistDiT]$^{z}$ & \textbf{891.53} & \textbf{21.43} & 0.4769 & \textbf{0.540} & \textbf{0.412} & \textbf{49.15} \\ \hline
Improvement to SoTA & -457.01 & 0.14 & -0.026 & 0.046 & -0.054 & -111.2 \\ \hline
\%age Improvement & \textcolor{green}{33.89\%} & \textcolor{green}{0.66\%} & \textcolor{red}{5.2\%} & \textcolor{green}{9.31\%} & \textcolor{green}{11.59\%} & \textcolor{green}{69.4\%} \\ \hline
\end{tabular}%
}
\end{table}
The quantitative results in Table ~\ref{tab:BCIcomparison}, demonstrate better improvement in terms of MSE, PSNR, FID, outperforming state-of-the-art GAN and diffusion-based methods on BCI data.
\begin{table}[b]
\caption{Level-wise comparison of HER-2 Biomarker generation on the BCI Dataset.}
\label{tab:her2_levels}
\centering
\resizebox{\textwidth}{!}{%
\begin{tabular}{c|c|c|c|c|c}
\hline
\textbf{HER-2 Scale} & \textbf{IQA Metrics} & \textbf{cycleGAN \cite{Xu}} & \textbf{Pyr. Pix2Pix\cite{LiuBCI}} & \textbf{cond. Diff \cite{Xuanhe2024}} & \textbf{HistDiT} \\ \hline

\multirow{3}{*}{\textbf{\begin{tabular}[c]{@{}c@{}}Level 0\\ \scriptsize{(38)}\end{tabular}}} 
 & PSNR $\uparrow$ & 15.92 & \underline{17.96} & 16.75 & \textbf{22.49} \\
 & SSIM $\uparrow$ & 0.3587 & 0.3758 & \underline{0.4595} & \textbf{0.5172}\\
 & LPIPS $\downarrow$ & 0.5210 & \underline{0.4683} & 0.5161 & \textbf{0.3945} \\ \hline

\multirow{3}{*}{\textbf{\begin{tabular}[c]{@{}c@{}}Level 1+\\ \scriptsize{(235)}\end{tabular}}} 
 & PSNR $\uparrow$ & 16.84 & \underline{19.92} & 15.89 & \textbf{23.22} \\
 & SSIM $\uparrow$ & 0.3712 & \underline{0.4099} & 0.4027 & \textbf{0.5046} \\
 & LPIPS $\downarrow$ & 0.5185 & \underline{0.4467} & 0.5790 & \textbf{0.3836} \\ \hline

\multirow{3}{*}{\textbf{\begin{tabular}[c]{@{}c@{}}Level 2+\\ \scriptsize{(446)}\end{tabular}}} 
 & PSNR $\uparrow$ & 16.21 & \underline{20.39} & 15.70 & \textbf{20.4} \\
 & SSIM $\uparrow$ & 0.3759 & \underline{0.3984} & 0.3746 & \textbf{0.4645} \\
 & LPIPS $\downarrow$ & 0.5025 & \underline{0.4524} & 0.5908 & \textbf{0.4376} \\ \hline

\multirow{3}{*}{\textbf{\begin{tabular}[c]{@{}c@{}}Level 3+\\ \scriptsize{(258)}\end{tabular}}} 
 & PSNR $\uparrow$ & 15.64 & \underline{18.23} & 15.41 & \textbf{18.33} \\
 & SSIM $\uparrow$ & 0.3583 & \underline{0.3859} & 0.3654 & \textbf{0.4052} \\
 & LPIPS $\downarrow$ & 0.5394 & \underline{0.5052} & 0.5887 & \textbf{0.4664} \\ \hline
\end{tabular}%
}
\end{table}
The significantly lower FID (49.15) indicates that our dual-stream architecture generates textures indistinguishable from real pathology. The PSNR of (21.43) sets a new benchmark for signal fidelity. Notably, while standard SSIM score is slightly lower than some pixel-aligned baselines (e.g., ASP) but SCM is superior. This discrepancy, highlights a limitation of the SSIM in masking biological structures. In Table~\ref{tab:her2_levels}, we separately evaluate image quality across different HER-2 expression levels (0, 1+, 2+, 3+) by splitting the BCI test dataset. Our proposed HistDiT demonstrates superior performance across all levels of HER2 biomarker, particularly the challenging level 2+ and 3+ categories, where it successfully preserves the structural fidelity (SCM) and perceptual quality required for accurate medical analysis.

\subsection{Experimental Results on MIST Dataset}
\begin{figure*}[htbp]
\centering
\setlength{\tabcolsep}{0.5pt} 
\renewcommand{\arraystretch}{0.4} 
\resizebox{\textwidth}{!}{%
\begin{tabular}{c c c c c c c c c}
\rotatebox{90}{\parbox{1.3cm}{\centering \textbf{H\&E}}} & 
\includegraphics[width=0.115\linewidth]{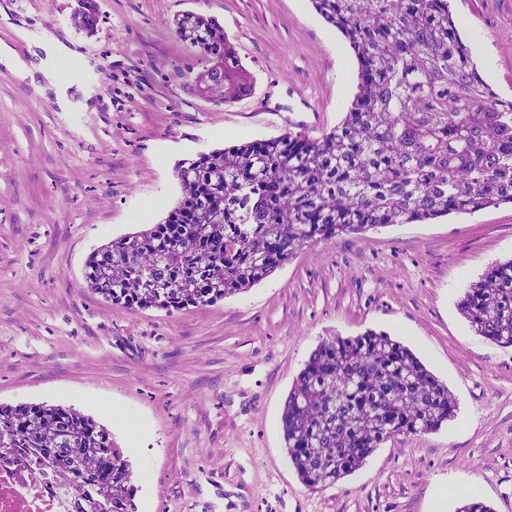} & 
\includegraphics[width=0.115\linewidth]{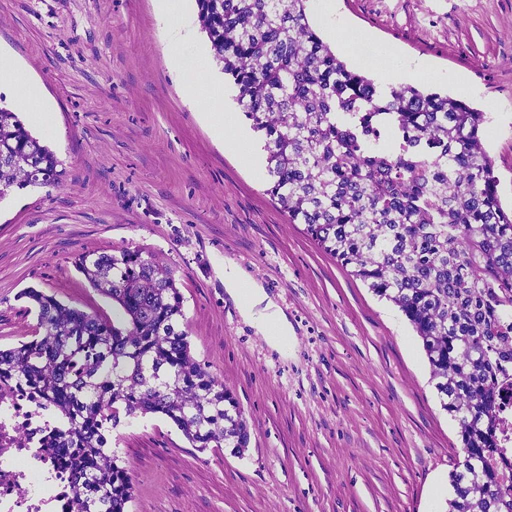} & 
\includegraphics[width=0.115\linewidth]{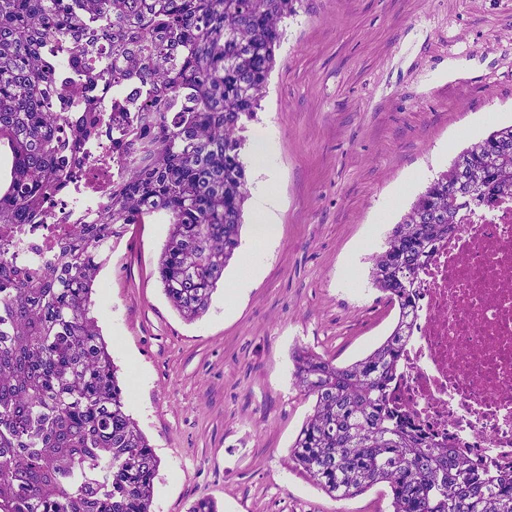} & 
\includegraphics[width=0.115\linewidth]{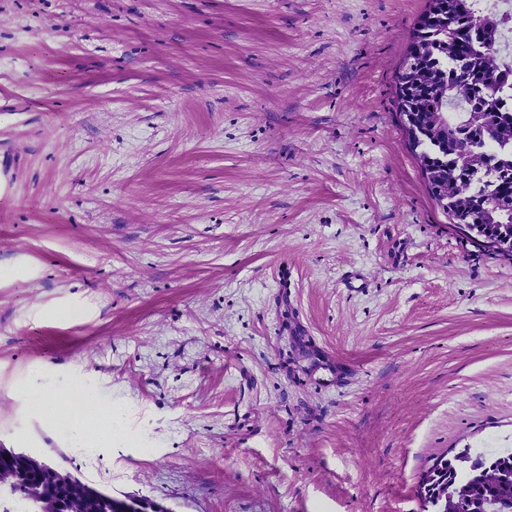} & 
\includegraphics[width=0.115\linewidth]{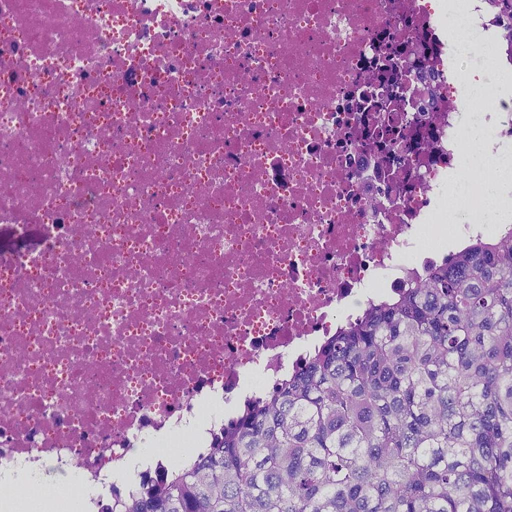} & 
\includegraphics[width=0.115\linewidth]{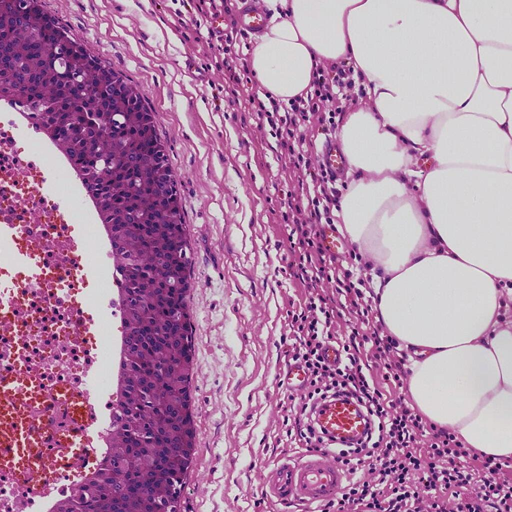} & 
\includegraphics[width=0.115\linewidth]{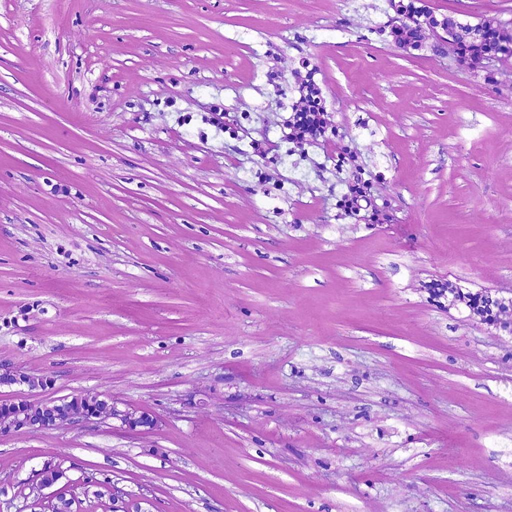} & 
\includegraphics[width=0.115\linewidth]{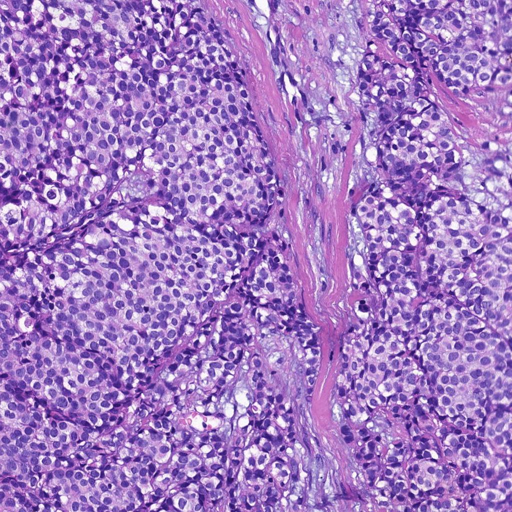} \\

\rotatebox{90}{\parbox{1.3cm}{\centering \textbf{cDiff}\cite{Xuanhe2024}}} & 
\includegraphics[width=0.115\linewidth]{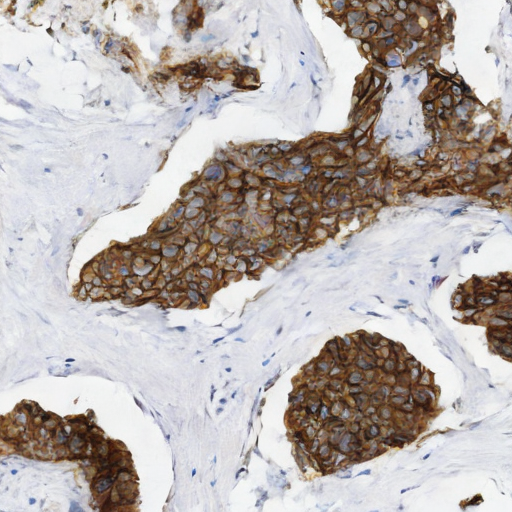} & 
\includegraphics[width=0.115\linewidth]{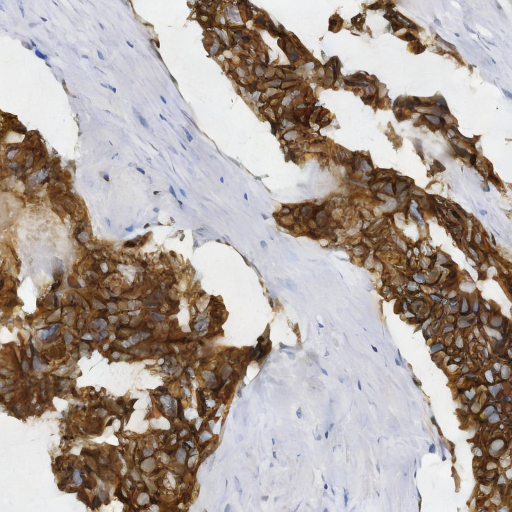} & 
\includegraphics[width=0.115\linewidth]{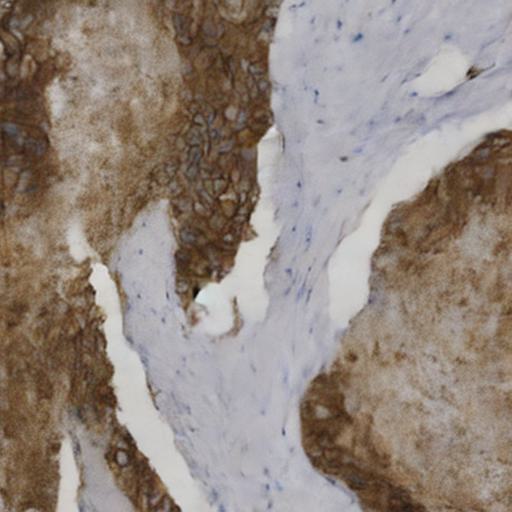} & 
\includegraphics[width=0.115\linewidth]{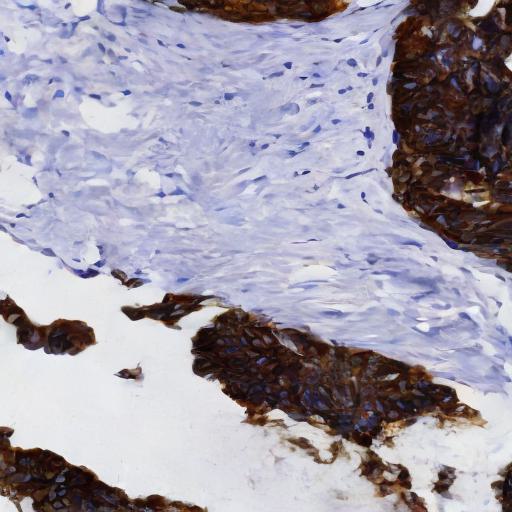} & 
\includegraphics[width=0.115\linewidth]{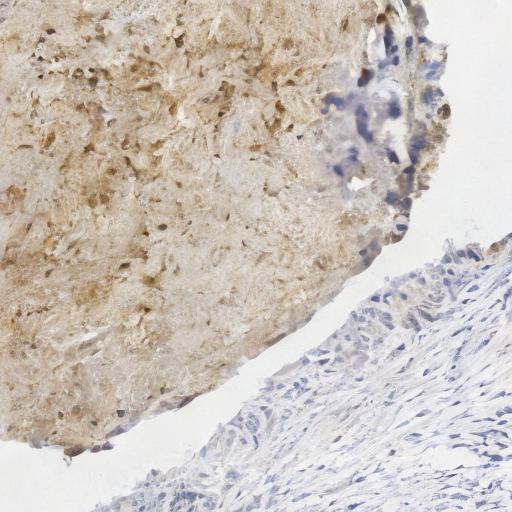} & 
\includegraphics[width=0.115\linewidth]{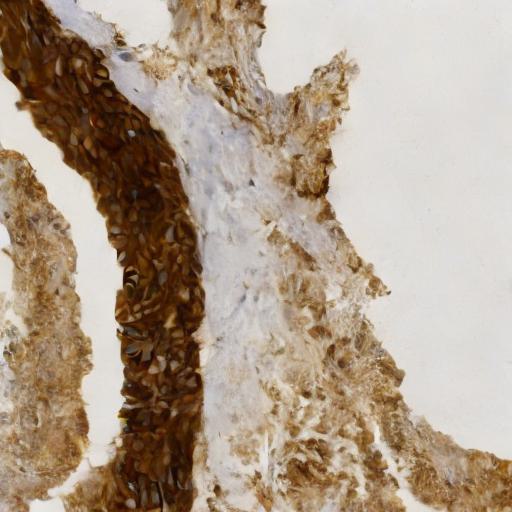} & 
\includegraphics[width=0.115\linewidth]{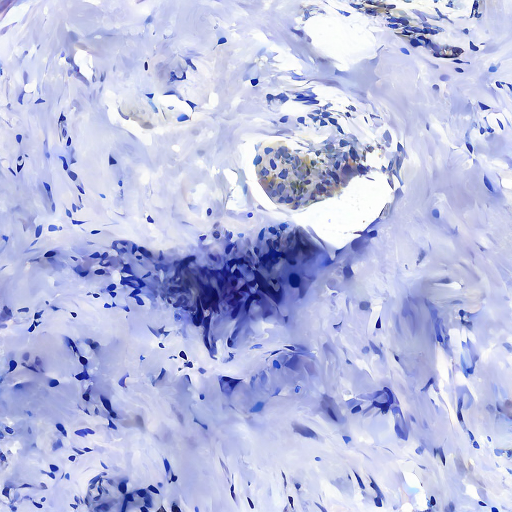} & 
\includegraphics[width=0.115\linewidth]{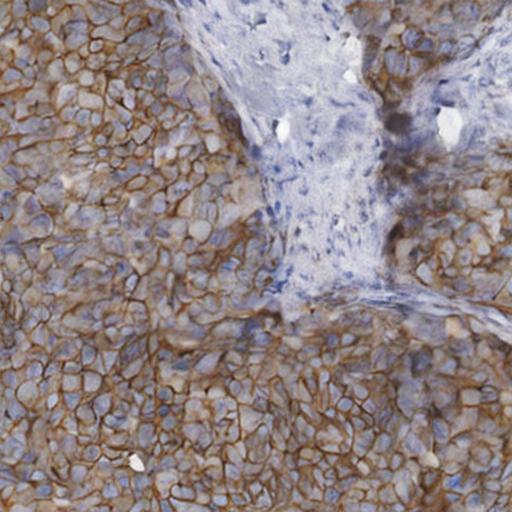} \\

\rotatebox{90}{\parbox{1.3cm}{\centering \textbf{pP2P}\cite{LiuBCI}}} & 
\includegraphics[width=0.115\linewidth]{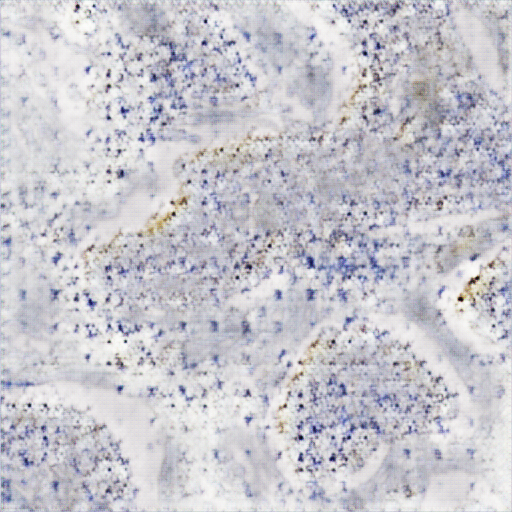} & 
\includegraphics[width=0.115\linewidth]{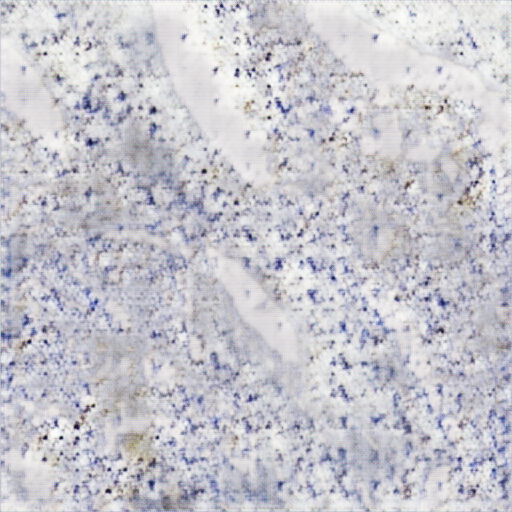} & 
\includegraphics[width=0.115\linewidth]{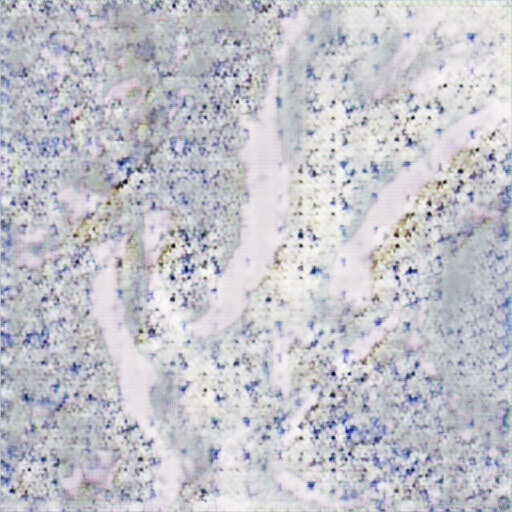} & 
\includegraphics[width=0.115\linewidth]{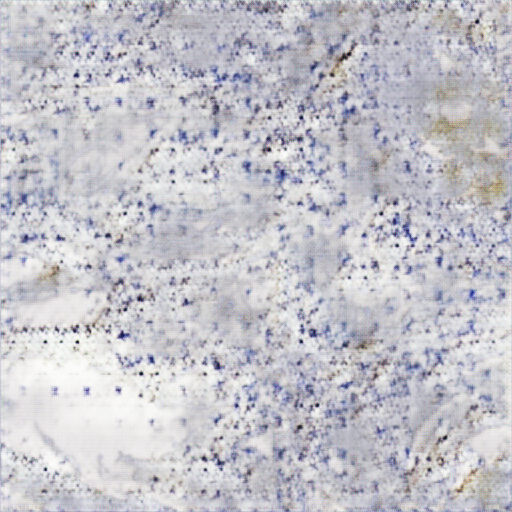} & 
\includegraphics[width=0.115\linewidth]{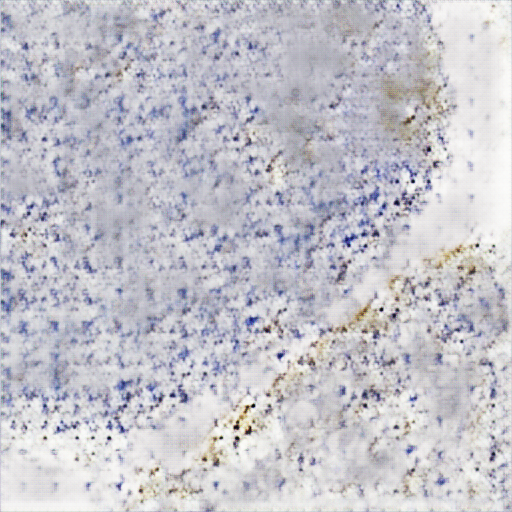} & 
\includegraphics[width=0.115\linewidth]{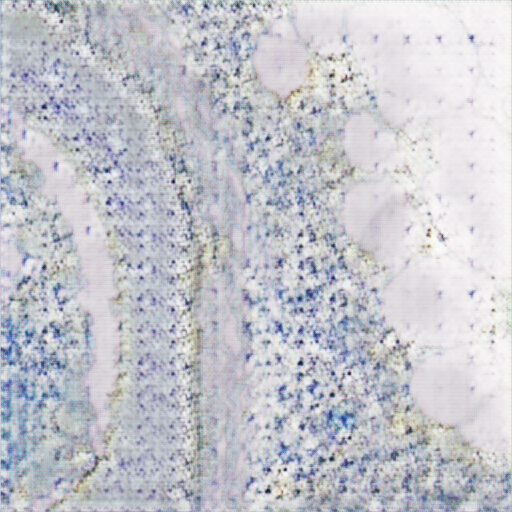} & 
\includegraphics[width=0.115\linewidth]{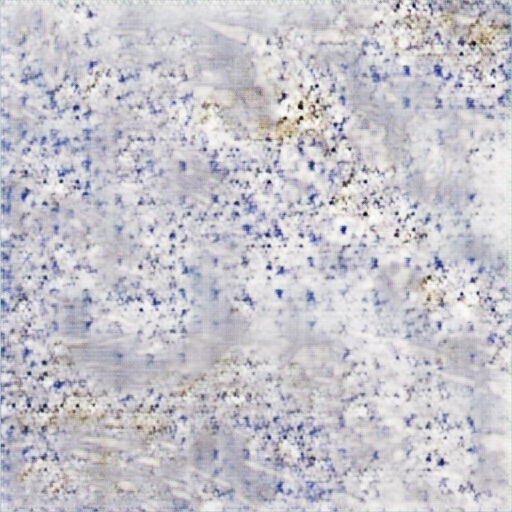} & 
\includegraphics[width=0.115\linewidth]{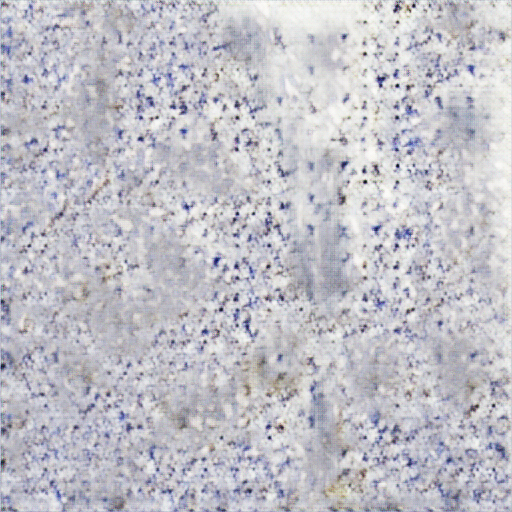} \\

\rotatebox{90}{\parbox{1.3cm}{\centering \textbf{Ours}}} & 
\includegraphics[width=0.115\linewidth]{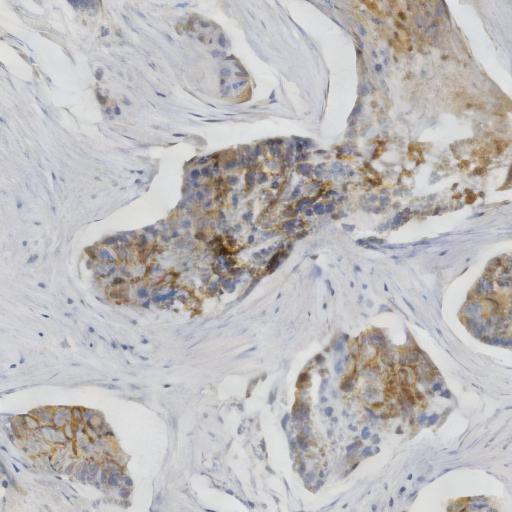} & 
\includegraphics[width=0.115\linewidth]{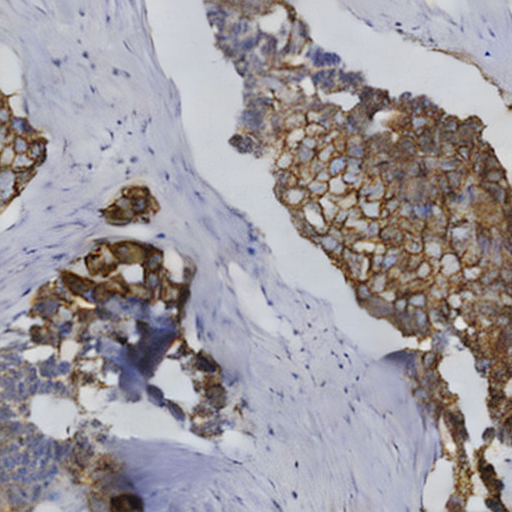} & 
\includegraphics[width=0.115\linewidth]{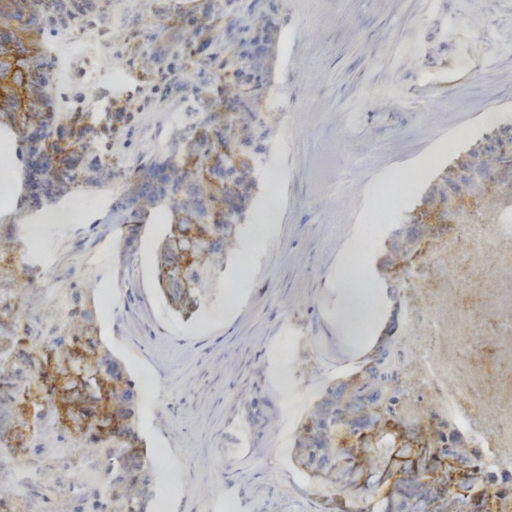} & 
\includegraphics[width=0.115\linewidth]{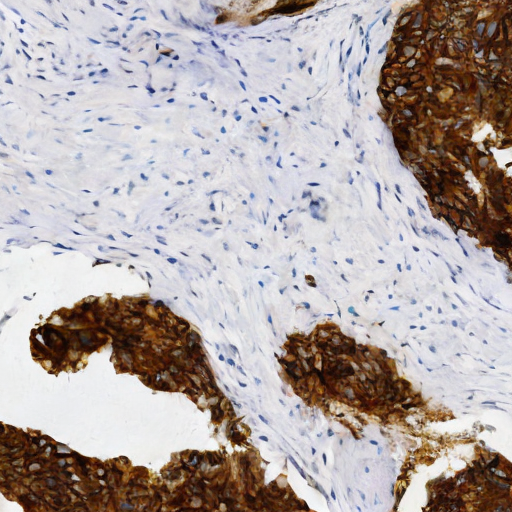} & 
\includegraphics[width=0.115\linewidth]{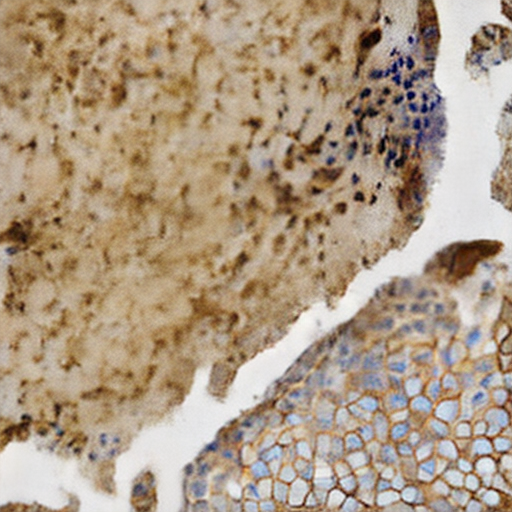} & 
\includegraphics[width=0.115\linewidth]{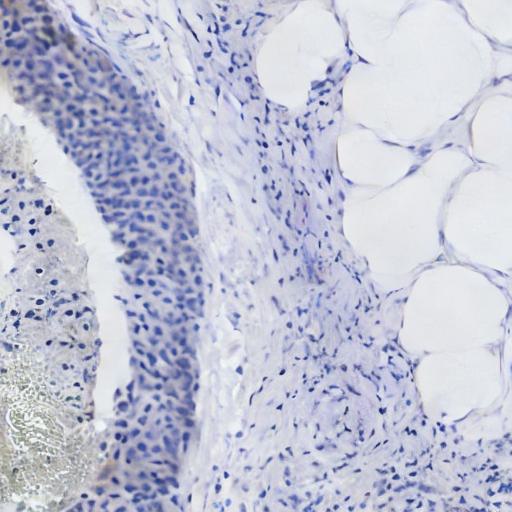} & 
\includegraphics[width=0.115\linewidth]{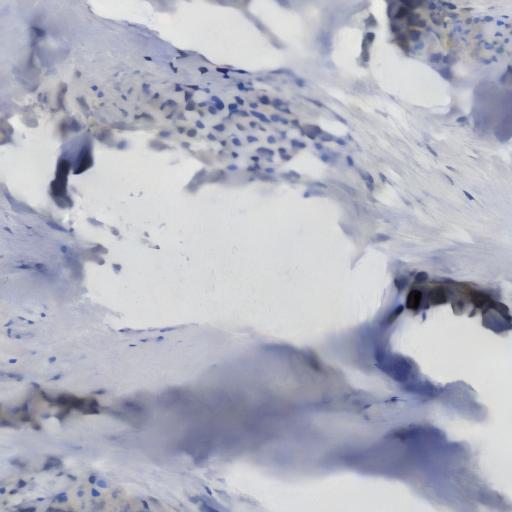} & 
\includegraphics[width=0.115\linewidth]{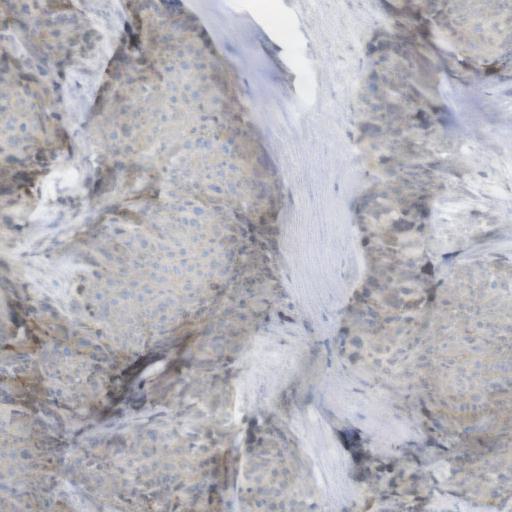} \\

\rotatebox{90}{\parbox{1.3cm}{\centering GT \textbf{IHC}}} & 
\includegraphics[width=0.115\linewidth]{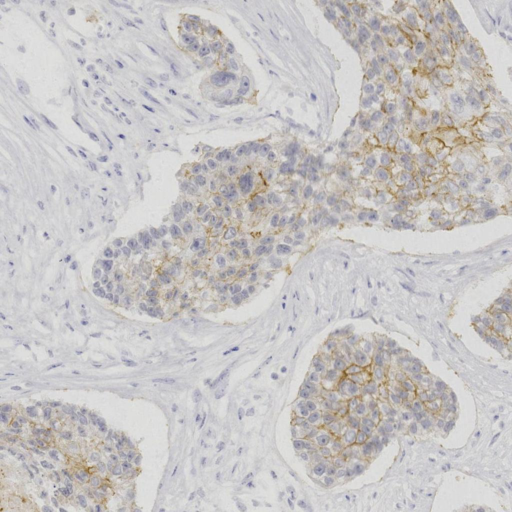} & 
\includegraphics[width=0.115\linewidth]{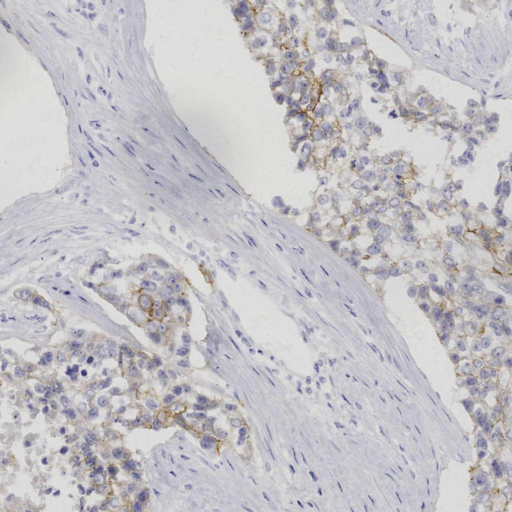} & 
\includegraphics[width=0.115\linewidth]{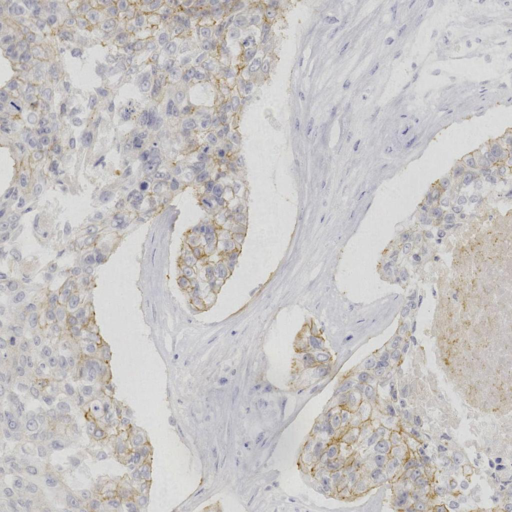} & 
\includegraphics[width=0.115\linewidth]{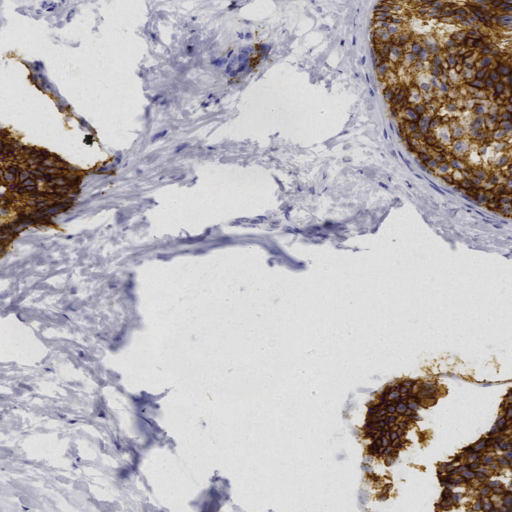} & 
\includegraphics[width=0.115\linewidth]{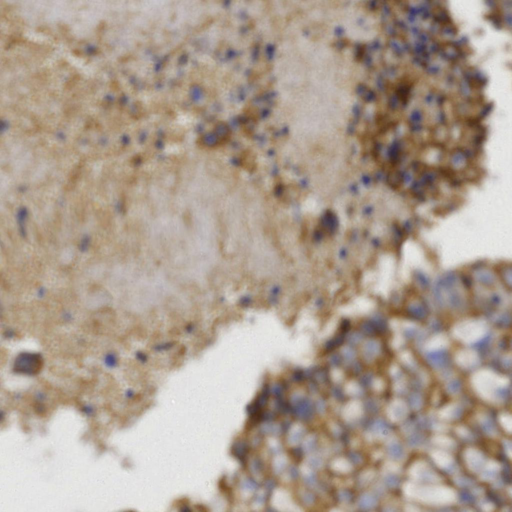} & 
\includegraphics[width=0.115\linewidth]{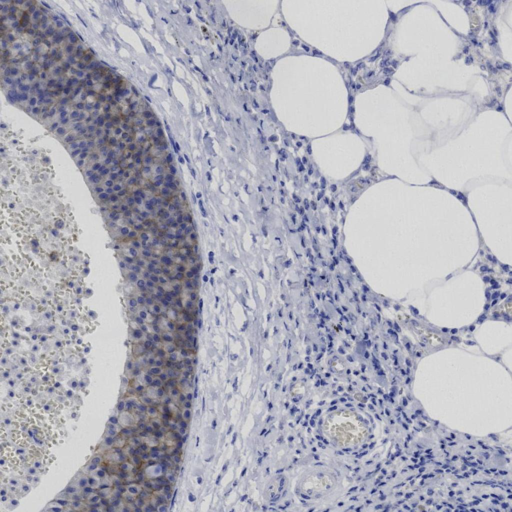} & 
\includegraphics[width=0.115\linewidth]{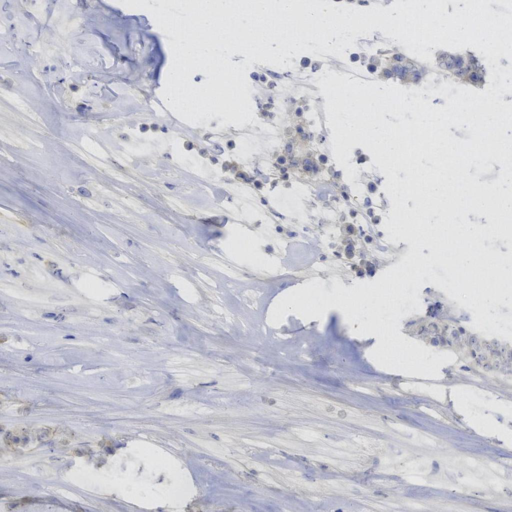} & 
\includegraphics[width=0.115\linewidth]{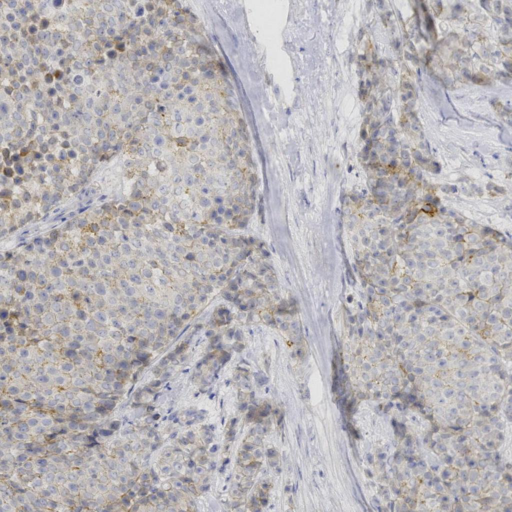} \\

\end{tabular} 
}
\caption{Visual comparison on the MIST dataset. Rows compare the input H\&E and baselines against our HistDiT and Ground Truth. HistDiT demonstrates superior stain restoration, generating sharp morphological details.}
\label{fig:mist_visuals}
\end{figure*}

The MIST dataset presents a significant challenge for generative models due to its inherent complexity and lack of standardized structural alignment. It is characterized by high intra-class variability and complex textural details that are difficult for conventional GANs to synthesize without artifacts. This makes it an ideal benchmark for testing a model's ability to maintain structural fidelity and perceptual realism under unconstrained conditions. In the test dataset we have 1000 input/output pairs for HER2 biomarker assessment. Visual quality analysis presented in Fig ~\ref{fig:mist_visuals}, reveals a superior staining quality of HistDiT.

\noindent We identified multiple instances, where the Ground Truth IHC images suffered from acquisition artifacts, like defocus blur or scanning noise. In these cases, HistDiT generated crispier, high-contrast images that were visually superior to the Ground Truth itself Fig. ~\ref{fig:objective_ablation}. By conditioning strictly on the sharp H\&E input, our model effectively restores the intended biological structures rather than over-fitting to the degraded quality of the target labels. Similarly, the results in Table~\ref{tab:mist_results} demonstrate that HistDiT achieves state-of-the-art performance across all reported metrics. Notably, HistDiT achieves SSIM of 0.211 and an FID of 59.3, outperforming competitors like HistDiST and PixCell. This indicates, when the data is noisy and unconstrained, our model's reliance on VAE structural blueprint provides a good stability that baseline approaches lack.
\begin{table}[htbp]
\caption{Quantitative comparison on the MIST Dataset. The best results are \textbf{bold} and second-best are \underline{underlined}. ($\uparrow$ higher is better, $\downarrow$ lower is better).}
\label{tab:mist_results}
\centering
\resizebox{\textwidth}{!}{%
\begin{tabular}{l|c|c|c|c|c|c}
\hline
\textbf{Method/Model} & \textbf{MSE}$\downarrow$ & \textbf{PSNR(dB)}$\uparrow$ & \textbf{SSIM}$\uparrow$ & \textbf{SCM}$\uparrow$ & \textbf{LPIPS}$\downarrow$ & \textbf{FID}$\downarrow$ \\ \hline

Cycle GAN \cite{Xu}$^{\rho}$ & 4190.52 & 11.91 & 0.174 & 0.209 & 0.553 & 125.7 \\
Pix2Pix \cite{Isola}$^{\rho}$ & 3485.20 & 12.74 & 0.150 & 0.194 & 0.614 & 128.1 \\
Pyramid Pix2Pix \cite{LiuBCI}$^{\rho}$ & \underline{2986.44} & \underline{14.24} & 0.165 & 0.214 & \underline{0.543} & 107.4 \\
ASP \cite{Fangda}$^{\rho}$ & --- & --- & 0.1945 & --- & --- & \textbf{54.28} \\
Conditional Diff. \cite{Xuanhe2024}$^{z}$ & 6252.98 & 11.13 & 0.168 & \underline{0.254} & 0.564 & 82.6 \\
HistDiST \cite{HistDiST}$^{z}$ & --- & --- & \underline{0.2059} & --- & --- & --- \\
PixCell (no LoRA) \cite{PixCell}$^{z}$ & --- & --- & 0.1880 & --- & --- & 67.68 \\ \hline

Proposed [HistDiT]$^{z}$ & 3396.88 & \textbf{14.26} & \textbf{0.211} & \textbf{0.302} & \textbf{0.489} & \underline{59.30} \\ \hline

Improvement to SoTA & -410.44 & 0.02 & 0.0051 & 0.048 & -0.054 & 5.02 \\ \hline
\%age Improvement & \textcolor{red}{13.74\%} & \textcolor{green}{0.14\%} & \textcolor{green}{2.5\%} & \textcolor{green}{18.9\%} & \textcolor{green}{9.94\%} & \textcolor{red}{9.25\%} \\ \hline
\end{tabular}%
}
\end{table}

\noindent These results suggest that proposed HistDiT learns the underlying concept of the stain distribution rather than merely memorizing pixel-to-pixel mappings, thus offering potential utility as a quality enhancement tool in clinical workflows.

\subsection{Expert Evaluation of Visual Fidelity}
To evaluate the perceptual realism and staining fidelity of our generated samples, we conducted a blind qualitative assessment with domain experts. The study involved immunologists from Cancer Research Malaysia (CRMY), who have extensive experience in IHC analysis. The experts were presented with anonymized random set of image patches containing both Real IHC (Ground Truth) and HistDiT-generated stains, paired with their corresponding H\&E. The experts were asked to distinguish the real biological samples from the virtually stained ones, serving as a visual Turing test. The feedback indicated that the experts were unable to consistently differentiate between the two, confirming that HistDiT successfully captures the complex staining patterns and structural coherence while minimizing the artifacts. While this validates the model's structural similarity and textural realism, we acknowledge that HER2 scoring is a specialized clinical task; therefore, strict diagnostic grading validation with board-certified pathologists is reserved for future clinical feasibility studies.

\subsection{Ablation Studies}
\begin{figure*}[htbp]
    \centering
    \begin{minipage}[b]{0.48\textwidth}
        \centering
        \resizebox{\linewidth}{!}{%
        \begin{tabular}{l|c|c|c}
            \hline
            \textbf{Method/Approach} & \textbf{PSNR}$\uparrow$ & \textbf{SSIM}$\uparrow$ & \textbf{FID}$\downarrow$ \\ 
            \hline
            $MSE$ only & 16.86 & 0.438 & 67.88 \\
            $L_1$ only & 19.44 & 0.450 & 93.31 \\
            $0.9 MSE + 0.1 L_1$ & 20.44 & 0.443 & 56.18 \\
            $0.7 MSE + 0.3 L_1$ & \textbf{21.43} & \textbf{0.477} & \textbf{49.15} \\ 
            \hline
        \end{tabular}
        }
        \vspace{0.2cm} 
    \end{minipage}
    \hfill
    \begin{minipage}[b]{0.5\textwidth}
        \centering
        \setlength{\tabcolsep}{0.7pt} 
        \renewcommand{\arraystretch}{0.5}
        \begin{tabular}{cccc}
            \includegraphics[width=0.23\linewidth]{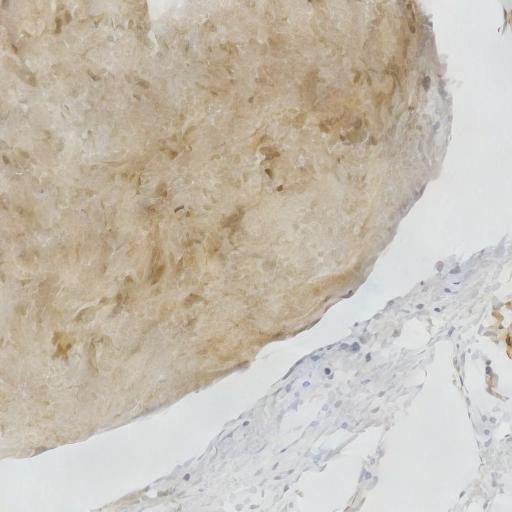} &
            \includegraphics[width=0.23\linewidth]{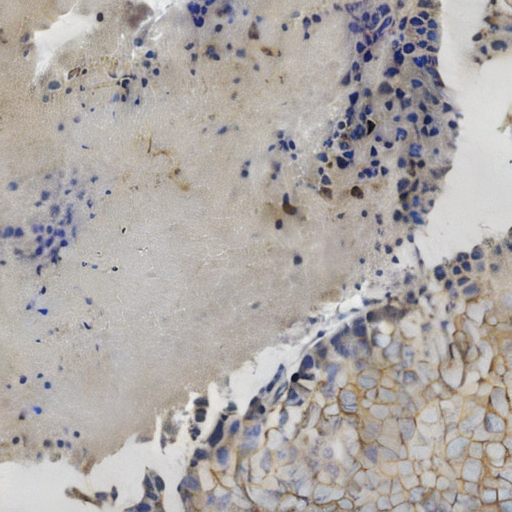} &
            \includegraphics[width=0.23\linewidth]{MISTResults/HistDiT/28M2101987_31_31.jpg_EMA.png} &
            \includegraphics[width=0.23\linewidth]{MISTResults/IHC_GT/28M2101987_31_31.jpg_IHC_GT.png} \\
            \\
            \scriptsize{\textbf{MSE only}} & 
            \scriptsize{\textbf{L1 only}} & 
            \scriptsize{\textbf{MSE+L1}} & 
            \scriptsize{\textbf{Real IHC}} \\
        \end{tabular}
        \vspace{0pt}
    \end{minipage}

    \caption{\textbf{Ablation Study on Objective Functions:} (\textit{Left}) Quantitative comparison of loss components on the BCI dataset. $MSE + L_1$ maintains high metric scores. (\textit{Right}) Visual samples demonstrating that the combined objective produces sharper structures (column 3) compared to the smoothing artifacts seen in MSE only (column 1).}
    \label{fig:objective_ablation}
\end{figure*}

To validate the effectiveness of our dual-stream conditioning and the hybrid objective function, we conducted a component-wise analysis. The architectural choices are detailed in Table~\ref{tab:Ablation1}, and impact of loss functions is shown in Fig.~\ref{fig:objective_ablation}.
\begin{table}[htbp]
\caption{Comparison for different architectural choices on BCI Dataset. Best are \textbf{bold}.}
\label{tab:Ablation1}
\centering
\resizebox{\textwidth}{!}{%
\begin{tabular}{l|c|c|c|c|c}
\hline
\textbf{Model Configuration} & \textbf{PSNR(dB)}$\uparrow$ & \textbf{SSIM}$\uparrow$ & \textbf{SCM}$\uparrow$ & \textbf{LPIPS}$\downarrow$ & \textbf{FID}$\downarrow$ \\ \hline

Pyramid Pix2Pix \cite{LiuBCI}$^{\rho}$ & 19.61 & 0.397 & 0.473 & 0.466 & 167.40 \\
HistDiT (Semantics only)$^{z}$ & 16.39 & 0.423 & 0.492 & 0.553 & 81.51 \\
HistDiT (Spatial only with Cross-Attention)$^{z}$ & 18.58 & 0.416 & \textbf{0.560} & 0.452 & 62.38 \\
HistDiT (Spatial [Concatenation], Semantics)$^{z}$ & 20.63 & 0.449 & 0.521 & 0.438 & 56.80 \\
HistDiT (Spatial [Cross-Attention], Semantics)$^{z}$ & \textbf{21.43} & \textbf{0.477} & 0.540 & \textbf{0.412} & \textbf{49.15} \\ \hline
\end{tabular}%
}
\end{table}

\noindent \textbf{Impact of Dual-Stream Conditioning:} 
Using UNI embeddings (Semantics Only) produced perceptually realistic textures but caused \textit{severe structural hallucinations}, generating high-density tissue artifacts. Conversely, removing the UNI guidance (Spatial Only) preserved tissue morphology but failed to resolve stain intensity between HER2 levels, often washing out dense expression levels. Furthermore, injecting spatial VAE-latents via cross-attention significantly outperformed standard channel-wise concatenation, confirming that dynamic attention is crucial for aligning structural blueprints with high-level semantic context.

\noindent \textbf{Optimization of Training Objectives:} 
We investigated the impact of training loss towards the reconstruction of high-frequency details (Fig.~\ref{fig:objective_ablation}). Training with ``MSE only'' resulted in over-smoothed textures, a known limitation where mean-square suppresses high-frequency variance. While the ``$L_1$'' loss improved structural contrast, it degraded visual fidelity (FID 93.31), likely due to the statistical mismatch between the Laplacian distribution of $L_1$ and the added Gaussian noise in forward diffusion. This forces the model to learn a suboptimal approximation of the noise distribution, degrading color fidelity. Our ``Hybrid Objective'' provides the optimal balance, utilizing $L_1$ to sharpen structures while retaining MSE for statistical consistency required for accurate stain-translation. The introduced hybrid objective helps in generating indistinguishable textures. 

\section{Conclusion}
This study introduces HistDiT, a dual-stream diffusion transformer for virtual immunohistochemistry. By integrating the structural precision of VAE-based spatial conditioning with the semantic richness of the UNI model, the architecture overcomes the limitations of previous GAN-based and U-Net diffusion approaches. The rigorous mathematical formulation of incorporating a scaled linear noise schedule and a robust hybrid loss function ensures training stability and visual fidelity. Additionally, the Structural Correlation Metric (SCM) rectifies the luminance bias of standard SSIM in histopathology. We compared HistDiT against established GAN and diffusion baselines on the BCI and MIST datasets, demonstrating superior performance across structural and perceptual quality metrics. Our findings confirm that HistDiT not only outperforms state-of-the-art methods but is capable of preserving diagnostic integrity required for clinical decision-making. This work positions DiTs, supported by domain-specific Foundation Models, as new standard for generative computational pathology. 

\subsubsection{Acknowledgements} This research is fully funded by the Edge Hill University's GTA studentship, and is in collaboration with Cancer Research Malaysia (CRMY) and University of Nottingham, Malaysia (UNM).

\end{document}